\documentclass{aa}

\usepackage{txfonts}
\usepackage{siunitx}
\usepackage{graphicx}   % Including figure files
\usepackage{algpseudocode}
\usepackage{booktabs}
\usepackage{hyperref}

\algnewcommand\Break{\textbf{break}}
\renewcommand{\part}{\text{part}}
\newcommand{\cell}{\text{cell}}
\newcommand{\SMBH}{\text{SMBH}}
\newcommand{\AGN}{\text{AGN}}
\newcommand{\Msun}{\ensuremath{M_{\odot}}}
\renewcommand{\tt}{\text{t}}
\renewcommand{\cell}{\text{cell}}
\renewcommand{\vec}[1]{{\bf #1}}

\usepackage{color}
\definecolor{Red}{rgb}{0.65,0.08,0.05}
\begin{document}

\title{Accurate tracer particles of baryon dynamics  \\ in the adaptive mesh refinement code Ramses}
\titlerunning{Tracer particles in Ramses}

\institute{
  Institut d'Astrophysique de Paris, CNRS \& UPMC, UMR 7095, 98 bis Boulevard Arago, 75014, Paris, France \label{IAP}\and
Korea Institute of Advanced Studies (KIAS) 85 Hoegiro, Dongdaemun-gu, 02455 Seoul, Republic of Korea\label{KIAS}
}
\author{
Corentin Cadiou\inst{\ref{IAP}} \and
Yohan Dubois\inst{\ref{IAP}} \and
Christophe Pichon\inst{\ref{IAP},\ref{KIAS}}
}

\date{Accepted for publication in A\&A}
\abstract{
We present a new implementation of the tracer particles algorithm based on a Monte Carlo approach for the Eulerian adaptive mesh refinement code {\sc Ramses}. The purpose of tracer particles is to keep track of where fluid elements originate in Eulerian mesh codes, so as to follow their Lagrangian trajectories and re-processing history.
We provide a comparison to the more commonly used velocity-based tracer particles, and show that the Monte Carlo approach reproduces the gas distribution much more accurately. 
We present a detailed statistical analysis of the properties of the distribution of tracer particles in the gas and report that it follows a Poisson law.
We extend these Monte Carlo gas tracer particles to tracer particles for the stars and black holes, so that they can exchange mass back and forth between themselves.
With such a scheme, we can follow  the full cycle of baryons, that is, from gas-forming stars to the release of mass back to the surrounding gas multiple times, or accretion of gas onto black holes. 
The overall impact on computation time is $\sim$ \SI{3}{\%} per tracer per initial cell.
As a proof of concept, we study an  astrophysical science case -- the dual accretion modes of galaxies at high redshifts --, which highlights how the scheme yields information hitherto unavailable. These tracer particles will allow us to study complex astrophysical systems where both efficiency of shock-capturing Godunov schemes and a Lagrangian follow-up of the fluid are required simultaneously.
}

%\usepackage{color}

%\label{firstpage}
% \pagerange{\pageref{firstpage}--\pageref{lastpage}}
\maketitle

\section{Introduction}

Many astrophysical problems of interest require us to solve equations of hydrodynamics on very different timescales and physical scales. Two main methods have been developed to solve these equations. On the one hand, one can study the motion of the gas by following the evolution of interacting particles. This Lagrangian approach is the one used by smooth particle hydrodynamics~\citep[SPH, e.g.][]{springel_cosmological_2005,wadsley_gasoline_2004,price_phantom_2017} codes. These codes sample the gas distribution using a set of fixed-mass macro-particles smoothed with a given kernel, and move particles accordingly. By construction, this approach provides the Lagrangian evolution of the gas. This property is also one of its shortcomings: low-density regions are populated by large particles and hence lack resolution.
On the other hand, gas hydrodynamics can also be described on a grid, where gas distribution is sampled on finite volumes, and solved with efficient shock-capturing Godunov solvers. Adaptive mesh refinement~\citep[AMR, e.g.][]{kravtsov_adaptive_1997,teyssier_cosmological_2002,springel_e_2010-1,bryan_enzo_2014}  codes follow this approach and allow for a dynamical refinement of the mesh. Though quasi-Lagrangian refinement is most commonly adopted in situations addressing galaxy formation problems, super-Lagrangian resolutions can also be achieved by refining the grid based on gas quantities such as the Jeans length to follow gravitationaly unstable star-forming regions~\citep{agertz_large_2009}, the vorticity to follow the seeding of turbulence~\citep[e.g.][]{ipapichino_hydrodynamical_2008}, the relative variation of any hydro quantity~\citep[such as e.g. the ionised fraction of hydrogen][]{rosdahl_extended_2012}, or using a passive scalar to keep track of a particular gas phase~\citep[such as for jets, see, e.g.][]{bourne_agn_2017}, among others. While super-Lagrangian refinement provides a very flexible method to trigger refinement, it falls short of providing the Lagrangian history of the gas.

To overcome this issue, AMR codes have been equipped with ``tracer'' particles. 
Tracer particles are passively displaced with the gas flow, and hence track its Lagrangian evolution. Each tracer can also be used to record instantaneous quantities, such as the thermodynamical properties of the gas or any other property. Many astrophysical problems can can benefit greatly from this Lagrangian information. For example, when studying galaxy formation, the past Lagrangian history of the gas is crucial to understand how gas has been accreted and how it has been ejected in large-scale galactic outflows. Tracer particles can be used to study the density and temperature evolution of the gas~\citep[e.g.][]{nelson_moving_2013,tillson_angular_2015} that will eventually form stars. For example, one could use tracer particles to study the temperature evolution of the gas as it falls onto galaxies, to study the number of dynamical times before it becomes star forming or to quantity the number of time gas is recycled in stars or sent in galactic fountains.% : \LEt{Please reform these questions into aims, if they are to be answered here. If attempts are not made to asnwer these questions here then these subjects could either be added at the end of the conclusions are suggestions for future work, or simply removed altogther. A list of such open questions may not help the reader here.}Did the gas heat-up as it fell onto the galaxy? How many dynamical times did it orbit the galaxy before forming a star? Did it come from a cold, entropy-poor region or from the diffuse hot halo? How many times is a parcel of gas recycled into a star? How long does it take for a parcel of gas to fall back onto a galaxy after a supernova event?
Another problem that requires the use of tracer particles is the study of mixing. Particularly in turbulent environments, such as the interstellar or the intergalactic medium, the Lagrangian information provides information about, for example, mixing timescale \citep[e.g.][]{federrath_turbulent_2008}, the origin of turbulence \citep[e.g.][]{vazza_massive_2011,vazza_turbulence_2012}, or how it contributes to core buildup \cite{mitchell_cores_2009}. In addition to this, the past Lagrangian evolution of a parcel of fluid can also impact the modelling itself \citep[e.g.][]{federrath_turbulent_2008,silvia_numerical_2010}.

In this paper we present a new implementation of tracer particles in the AMR {\sc Ramses} code~\citep{teyssier_cosmological_2002}. This implementation is based on the one developed by~\cite{genel_following_2013} for the moving mesh {\sc arepo} code~\citep{springel_e_2010-1}. It has been extended to track the full Lagrangian history of baryons in any phase, including their conversion from gas to stars, from stars back into the gas {\em via} supernova feedback, their interaction with feedback from black holes, and their accretion onto them. This ``Monte Carlo'' (MC) tracer particle implementation improves the previous implementation, ``velocity''-advected tracers. With the velocity-based approach, tracer particles are moved based on the interpolated local values of the gas velocity field. While this yields qualitative results, it suffers from systematic effects: tracer particles over-condensate in regions of converging flows~\citep{genel_following_2013}. Monte Carlo tracer particles follow a different idea. They are moved so that the tracer particle mass flux at each cell interface is statistically equal to that of the gas. Thanks to this property, the Eulerian distribution of tracers converge to that of the gas when the number of tracer particles goes to infinity. In addition to matching the gas distribution, the implementation of tracer particles here is also able to match the distribution of baryons in stars and in black holes.

The paper is structured as follows.  Section~\ref{sec:implementation} details the implemented algorithm.  Section~\ref{sec:tests} presents tests and validations of the new implementation. In particular,  Section~\ref{sec:idealized-tests} presents the results from idealised tests and Section~\ref{sec:astrophysical-tests} presents an analysis of the properties of  tracers in a real astrophysical simulation. 
Using the same simulation, Section~\ref{sec:science_case} illustrates the efficiency of the scheme applied to a specific science case -- the bimodal accretion of gas onto galaxies at high redshift. 
Section~\ref{sec:performance} assesses the performance of the scheme.
Section~\ref{sec:conclusions} provides a discussion of our results and our  conclusions.
Appendix~\ref{sec:algo} provides more details about the algorithm.

\section{Implementation}%
\label{sec:implementation}

The {\sc Ramses} code~\citep{teyssier_cosmological_2002} solves the full set of Euler equations by formulating the equations in terms of finite-volume, that is, by calculating fluxes at the interfaces of cells of the adaptive mesh.
This is done by using a MUSCL-Hancock method with a second-order Godunov solver calculating the fluxes from linearly interpolated values at cell faces from the cell-centred values limited by a total-variation-diminishing scheme. 
Such a Eulerian-based method has proven efficient at capturing shock discontinuities and achieves efficient mixing of shear layers of gas; however, its main drawback is that it does not naturally provide the Lagrangian trajectories of gas elements. 

To address this problem, it is possible to introduce the so-called tracer particles of the flow that should follow the flow lines of the gas.
A naive approach to track the motion of the gas is to use the velocity of the gas itself, assign it to tracer particles, and move them accordingly. 
This is done with a cloud-in-cell interpolation of the velocity values of the overlapped cells where the volume of the cloud is that of the host cell, though the level of the interpolation is not particularly important~\citep[nearest grid point or triangular shape cloud;][]{federrath_turbulent_2008}.
Such a velocity-based approach was implemented in {\sc Ramses}~\citep{dubois_feeding_2012} and used to probe the link between cosmic gas infall and galactic gas feeding, and its acquisition of angular momentum~\citep{pichon_rigging_2011,dubois_feeding_2012,tillson_angular_2015}.
While this approach yields smooth trajectories, it falls short of reproducing the gas density distribution accurately in regions with strong convergence of the velocity field~\citep{genel_following_2013}.

To address this shortcoming, we have implemented in {\sc Ramses} the MC approach of tracer particles introduced by~\cite{genel_following_2013} for {\sc arepo}~\citep{springel_e_2010-1}. Instead of having proper velocities and positions, MC tracers are attached to individual cells and are allowed to `jump' from the centre of one cell to the centre of another according to the mass fluxes obtained through the Godunov solver. 

\begin{figure}
  \centering
  \includegraphics[width=\columnwidth]{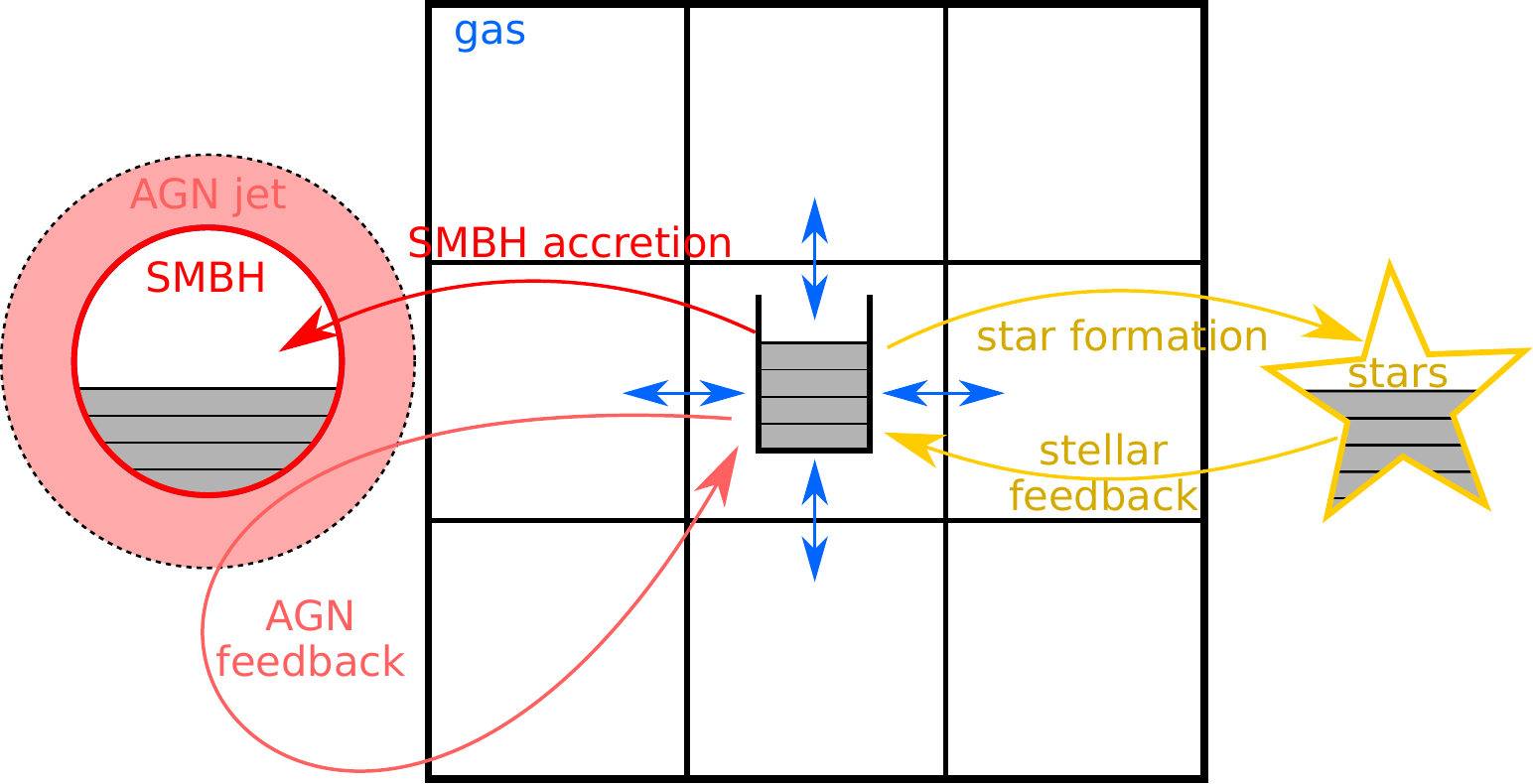}
  \caption{Scheme of the different ``buckets'' that can hold tracer particles and the process that moves them around. The three buckets are gas cells, stars, and SMBHs. Arrows indicate outgoing mass fluxes between buckets and the physical process associated, and grey squares represent tracer particles. The jet mode feedback from AGNs (around SMBHs) is able to move gas tracer particles from the central cell to the surrounding cells. The particles have no spatial distribution within the buckets or any phase-space distribution. Tracer particles are exchanged probabilistically between buckets based on the mass fluxes. For example,\ for the gas, they are exchanged based on the mass fluxes at the boundary of the cells.}\label{fig:scheme-tracers}
\end{figure}

We have generalised the MC method to track exchanges of baryons between gas, star particles, and supermassive black hole (SMBH) particles, and in the following we refer to them as ``buckets''. At each time step, tracers are allowed to jump from any bucket $i$ to any bucket $j$ with a probability (gas$\rightarrow$gas, gas$\leftrightarrow$star, gas$\rightarrow$black hole) of
\begin{equation}
  \label{eq:base-proba}
  p_{ij} = \left\{\begin{matrix}
    \displaystyle \frac{\Delta M_{ij}}{M_i}\,, & \text{if $\Delta M_{ij} \geq 0$},\\
    0\,, & \text{if $\Delta M_{ij} < 0$},
  \end{matrix}\right.
\end{equation}
where $\Delta M_{ij}$ is the mass flowing from bucket $i$ to bucket $j$ between $t$ and $t+\Delta t$ and $M_i$ is the mass of the depleted bucket $i$ at time $t$. This probability is also the fraction of baryons flowing from one bucket to another. If the initial Eulerian distributions of tracers and baryons are equal, then in the limit where the number of tracers becomes large, satisfying Eq.~\eqref{eq:base-proba} is sufficient for the Eulerian distributions to remain equal at all times. Here is an outline of the proof. For any bucket $i$ containing $N_\tt$ tracers of equal mass $m_\tt$, let the total tracer mass read $M_\tt \equiv N_\tt m_\tt$. Because tracers are moved stochastically, the tracer mass flux $\Delta M_{\tt,ij}$ is a random variable. If at time $t$, $M_\tt = M_i$ (i.e. the Eulerian distributions are the same), then the {expected} tracer flux is $\mathrm{E}\left[\Delta M_{\tt,ij}\right] = N_\tt\times p_{ij} m_\tt = M_i p_{ij} = \Delta M_{ij}$. When the number of tracers becomes large, the tracer mass flux converges to the baryon flux, $\Delta M_{\tt,ij}\rightarrow \Delta M_{ij}$. The buckets have the same initial mass and are updated with the same mass fluxes, so they remain equal at the next time step, $t+\Delta t$. Therefore, if the initial Eulerian distributions are equal, by induction they remain equal at all times (in the limit of a large number of tracers)\footnote{In general, any stochastic scheme for which the expected tracer flux equals that of the baryons is able to track the Eulerian distribution at all times.}.

All the processes that are able to move tracers from bucket to bucket are summarised in Fig.~\ref{fig:scheme-tracers}. Tracers can move from one gas cell to another through gas dynamics, and the jet mode of AGN feedback from SMBHs, from gas to stars  via star formation, from stars to gas {via} supernova (SN) feedback, and from gas to SMBHs {via} black hole accretion. Below, we present the different algorithms used for each of these processes.

\subsection{Gas dynamics}

The algorithm moving tracer particles from one gas cell to another is the following. For each level of refinement, all the unrefined leaf cells are iterated over. For each leaf cell $i$ containing tracer particles, the total {outgoing} mass is computed as $\Delta M \equiv \sum_{j=1}^{2N_{d}} \max(\Delta M_{ij}, 0)$, where $j$ runs over the index of the neighbouring cells, $N_{d}$ is the number of dimensions, and $\Delta M_{ij}$ is the mass transferred between cell $i$ and cell $j$ in one time step and obtained from the Godunov flux of mass $F_{m,\rm ij}$, that is, $\Delta M_{\rm ij}=F_{m,\rm ij}\Delta t$. We take
\begin{equation}
    p_\mathrm{gas} = \frac{\Delta M}{M_i}
\end{equation}
to be the probability of displacing a gas tracer particle from one cell to any other of its neighbouring cell, and 
\begin{equation}
    p_{j} = \max\left(\frac{\Delta M_{ij}}{\Delta M}, 0\right)
\end{equation}
to be the probability of moving this tracer particle into cell $j$. For each tracer particle in cell $i$, a random number $r$ is drawn from a uniform distribution {between 0 and 1}. If $r< p_\mathrm{gas}$, the tracer is selected. For each selected tracer, another random number $r'$ is drawn. For each neighbouring cell $j$ with a positive flux (such that $\Delta M_{ij} > 0$), if $r'<p_j$ the tracer particle is moved into cell $j$ and the algorithm proceeds to the next particle; else, $r'$ is decreased so that $r' \leftarrow r'- p_j$ and the algorithm proceeds to the next neighbouring cell. Because the sum of all the $p_j$ is 1, this procedure will assign the tracer to a single cell.

When a cell of mass $M_0$ is refined between two time steps, all its tracers are distributed randomly to one of the eight newly created cells, the probability for a tracer particle to be attached to the new cell $i$ being $p=M_i/M_0$ (refined density can be interpolated from neighbouring values or equally distributed amongst new cells). Conversely when a cell is derefined all its tracers are attached to the parent cell.

\subsection{Star formation}

This part of the algorithm involves moving tracers from the gas phase into star particles, and moving the star-tracer particles along with their star particles.

We first recall that the star formation process in {\sc Ramses} is usually modelled by a Schmidt law, where the star formation rate density is non-zero and 
\begin{equation}
    \frac{d \rho_\star}{d t}=\epsilon_\star \frac{\rho_{\rm g}}{t_{\rm ff}}\, , 
\end{equation}
when $\rho_{\rm g}>\rho_0$, and where $\rho_{\rm g}$ is the gas density, $\rho_0$ a gas density threshold, $t_{\rm ff}=(3\pi/(32G\rho_{\rm g}))^{1/2}$ the gas free-fall time, and $\epsilon_\star$ the efficiency of star formation, which can be taken as an \emph{ad hoc} constant, or as a function of the local gravo-turbulent properties of the gas~\citep{krumholz_general_2005,hennebelle_analytical_2011,padoan_star_2011}.
A single star particle made of $N_\star$ stars of mass resolution $M_{\star,0}$ is created, where $N_\star$ is drawn according to random Poisson process~\citep{rasera_history_2006}:
\begin{equation}
    P_{\rm sf}=\frac{\lambda^{N_\star}}{N_\star!}\exp{(-\lambda)}\, ,
\end{equation}
where $P_{\rm sf}$ is the probability of creating $N_\star$ particles from the gas (and accordingly removing $M_\star\equiv N_\star M_{\star,0}$ mass from the gas cell), and 
\begin{equation}
    \lambda=\frac{\rho_{\rm g}\Delta x^3}{M_{\star,0}}\frac{\Delta t}{\epsilon_\star^{-1}t_{\rm ff}}\, .
\end{equation}

Finally, the transfer of gas tracer particles to star-tracer particles at time of creation $t$ of $M_\star$ is given by the probability
\begin{equation}
    p_\star = \frac{M_\star}{M_i}.
\end{equation}
In more details, we loop over all the gas tracer particles contained in the cell where the new star is created. For each of them, a random number $r$ is drawn from a uniform distribution {between 0 and 1}. If $r<p_\star$, the gas tracer particle is turned into a star-tracer particle at the same position as that of the star particle (i.e.\ at the centre of the cell). The star-tracer particle is `attached' to the star particle by moving along with the star particle, which is done through a classic leap-frog integration of its motion. Therefore, at all time steps, the position of the tracer is updated to match the position of its star. The index of the star is also recorded on the tracer for convenience.

The implementation also comes with two alternatives to initialise the tracer particles. One can feed in a list of positions to the code; 
one tracer will be created at each location. Alternatively, we developed an initialisation scheme that takes as input the mass that each tracer particle represents, $m_\tt$. The scheme is called ``in-place initialisation'' as it is performed directly within the code: the scheme is called once at startup, after the AMR grid has been built. It loops over all cells, and for each of them computes the number of tracer particles to create. The expected number of tracers created in a cell of mass $M_\cell$ is $N= m_\tt/M_\cell$. Let us write $N_0 = \lfloor N \rfloor$. The scheme creates $N_0\equiv \lfloor N \rfloor $ particles in the cell and then creates an additional one with probability $N-N_0$. In the end, the expected number of tracer particles created in the cell is $N$, meaning that on average each cell is populated with the correct number of tracer particles. In the following, unless stated otherwise, the tracer particle distribution is always initialised using the in-place method.

\subsection{Supernova feedback}
%%%%%%%%%%%%%%%%%%%%%%%%
\begin{figure}
  \centering
  \includegraphics[width=.5\columnwidth]{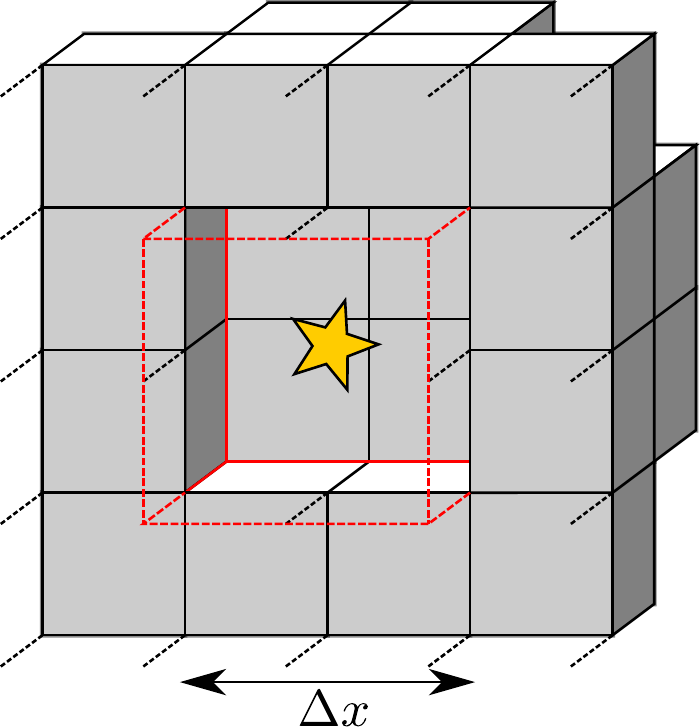}
  \caption{Scheme of the 48 neighbouring virtual cells (only the 24 rear ones are shown) where mass and momentum are deposed during a SN event. The cell containing the SN has a size of $\mathrm{\Delta} x$ and is outlined in red.}\label{fig:scheme-SN-feedback}
\end{figure}

Let us describe the transfer of mass of a star particle to the gas according to type II SN explosions (referred to henceforth as SNII) and their associated tracer particles. This can be trivially extended to the more complete description of the evolution of stellar mass loss.

When a star particle sampling an initial mass function (IMF) of mass $M_\star$ explodes into type II SNe, it releases a mass $\eta_{\rm SN} M_\star$, where $\eta_{\rm SN}$ can be crudely approximated by the mass fraction of the IMF going SNII. The probability of releasing a star-tracer particle into the gas is $p_\mathrm{SN} = \eta_{\rm SN}$.
For each star particle turning into SNe, we loop over all the star-tracer particles attached to it. For each of these, a random number $r$ is drawn from a uniform distribution {between 0 and 1}. If $r<p_\mathrm{SN}$, the star-tracer particle is released in the gas and turned into a gas tracer particle. Otherwise, the tracer is left attached to the stellar remnant.

The transfer of star-tracer particles to the gas by SNII is described here for the so-called mechanical feedback model of~\cite{kimm_escape_2014}~\citep[see also][]{kimm_towards_2015}\footnote{We have extended this implementation to i) simple thermal pulses of energy (with or without delayed cooling~\citealp{teyssier_cusp_2013}), where the mass is released to the central cell only, and ii) to the so-called kinetic model of~\cite{dubois_onset_2008} (in its more recent form described in~\citealp{rosdahl_snap_2017}) where ``debris'' particles are replaced by a bubble injection region of energy, momentum, and mass according to the Sedov-Taylor solution.}. In this model, the gas is released into the neighbouring cells. 
The mechanical feedback scheme is designed to overcome the consequences of radiative losses in SN bubbles due to the lack of resolution.
Where the cooling time of the SN-heated gas is shorter than the hydrodynamical time step, the energy-conserving phase (with Sedov-Taylor solution), during which the momentum is growing by the pressure work of the bubble, cannot be captured properly, and leads to spurious energy {and} momentum loss. 
To circumvent this problem, \cite{kimm_escape_2014} introduced a model that correctly accounts for the momentum injection according to the Sedov-Taylor or snow-plough solution~\citep{thorton_energy_1998}, which depends on the cooling rate of the gas, or more precisely on the energy release, local gas density, and metallicity.
The cell containing the exploding star particle is virtually represented by 8 cells refined by an additional level, which are equivalently surrounded by 48 such virtual neighbouring cells, as illustrated in Fig.~\ref{fig:scheme-SN-feedback}~\citep{kimm_escape_2014}. 
The mass ejecta together with the mass of the swept-up gas of the central true cell is released evenly in all the virtual cells, and is attributed back accordingly to the true existing cells. 

The tracer particles are interfaced with this feedback model as follows: For each released star to gas tracer particle, a random integer number $l\in[1,48]$ is drawn uniformly. The star tracer is then moved to the centre of the $l$-th virtual cell and attributed to the corresponding true existing cell as a new gas tracer particle.

\subsection{SMBH formation and gas accretion}

Supermassive black holes are modelled as sink particles that can form out of the dense star-forming gas, grow by accretion of gas, and coalesce following the implementation described in~\cite{dubois_self-regulated_2012}.

A SMBH forms according to several user-defined criteria, typically above a given gas density threshold $\rho_0$ and outside an exclusion distance radius $r_{\rm ex}$ within which SMBH is artificially prevented if any other SMBH already exists (in order to prevent creation of multiple SMBHs within the same galaxy).
When a SMBH forms with an initial seed mass $M_{\rm SMBH,0}$, gas tracer particles in the cell of mass $M_i$ containing the SMBH are attached to the SMBH and become SMBH tracer particles according to a probability
\begin{equation}
  p_\SMBH = \frac{M_{\rm SMBH,0}}{M_i}.
\end{equation}

SMBHs can continuously accrete gas according to the Bondi-Hoyle-Littleton accretion rate, capped at Eddington with
\begin{align}
  \dot{M}_\SMBH & = (1-\varepsilon_r)\ \dot{M}_\text{acc} = (1-\varepsilon_r)\ \text{min}(\dot{M}_\text{B}, \dot{M}_\text{Edd}),\\
  \dot{M}_\text{B} & = \frac{4\pi \rho G^2 M_\SMBH^2}{(c_s^2 + u^2)^{3/2}} {\left(\frac{\rho}{\rho_\mathrm{boost}}\right)}^\alpha,\\
  \dot{M}_\text{Edd} & = \frac{4\pi G m_p M_\SMBH}{\sigma_T\varepsilon_r c},
\end{align}
where $\dot{M}_{\rm acc}$, $\dot{M}_\SMBH$, $\dot{M}_\text{B}$, and $\dot{M}_\text{Edd}$ are the disc, SMBH, Bondi-Hoyle-Littleton, and Eddington accretion rates, respectively, $m_p$ is the mass of a proton, $G$ the gravitational constant, $\sigma_T$ the Thomson cross-section, $\varepsilon_r$ the radiative efficiency, $c_s$ the speed sound, $u$ the mean velocity of the gas with respect to\ the SMBH, and $c$ the speed of light. $\rho_\text{boost}$ and $\alpha$ are free parameters set, here, to $\rho_\text{boost} = \SI{8}{m_p.cm^{-3}}$ and $\alpha=2$ introduced to boost the accretion rate due to unresolved small-scale larger densities~\citep{booth_cosmological_2009}. The value of $\varepsilon_r$ is either chosen as a constant value equal to $0.1$, or, here, as a varying function of the spin of SMBH, whose evolution is governed by gas accretion and BH coalescence (see \citealp{dubois_black3_2014,dubois_black2_2014}, and Dubois et al., in prep.,\ for details). 

The mass taken from the gas cell in one time step is $\Delta M_\text{acc} \equiv \Delta t\, \text{min}(\dot{M}_\text{BH}, \dot{M}_\text{Edd})$. We note that $\Delta M_\text{acc} > \dot{M}_\SMBH\Delta t$ as part of the accreted mass is radiated away due to relativistic effect (and lost to the simulation). Each gas tracer in the cell containing the SMBH at a given time is accreted into the black hole with a probability of
\begin{equation}
  p_{\text{acc}} = \frac{\Delta M_\text{acc}}{M_i}.
\end{equation}
Tracer particles also model SMBH merger events. All the tracer particles attached to the two parent SMBHs are moved to the newly formed SMBH. There is no mechanism to extract tracers from the SMBH (reflecting the fact that there is no way to extract matter from a BH). One should also note that the total mass of SMBH tracers is larger than the total mass of SMBHs, as part of the energy-mass has been radiated away during accretion (and tracers have a fixed mass).

\subsection{AGN feedback}%
\label{sec:agn-implem}

\begin{figure}
  \centering
  \includegraphics[width=.6\columnwidth]{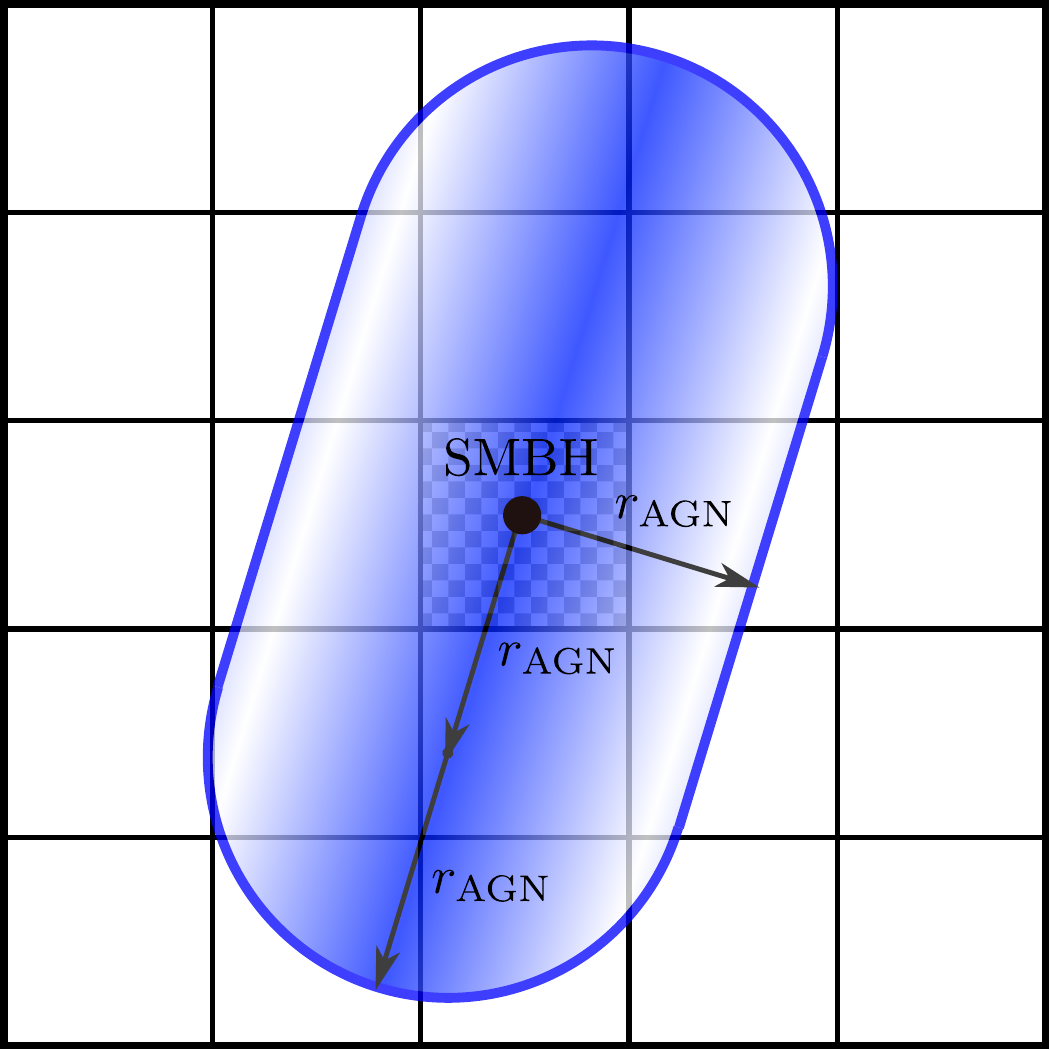}
  \caption{Schematic representation of the jet model. Gas is transported from the central cell (hatched region) containing the SMBH (black dot) into the jet (blue shaded region). The radial profile of the jet is a Gaussian of scale $r_\text{AGN}$. The shape of the jet is a ``capsule'' (a cylinder capped with two half spheres).}\label{fig:jet-scheme}
\end{figure}

In~\cite{dubois_self-regulated_2012}, there are two modes of AGN feedback: a quasar/heating mode and a radio/jet mode. The mode is selected based on the ratio of the Bondi-Hoyle-Littleton accretion rate to the Eddington accretion rate $\chi=\dot{M}_\text{B}/\dot{M}_\text{Edd}$. If $\chi<0.01$, the AGN is in jet mode, and, otherwise, it is in quasar mode~\citep{merloni_synthesis_2008}. 

In quasar mode, all the energy of the AGN proportional to $E_{\AGN, Q}=\varepsilon_{\rm f,Q}\varepsilon_r\dot{M}_{\rm acc}c^2\Delta t$ (the value $\varepsilon_{\rm f,Q}=0.15$ is calibrated to match the BH-to-galaxy mass relation~\citealp{dubois_self-regulated_2012}) is released as thermal energy in all cells within a sphere of size $R_\AGN$ and the mass of the gas is left untouched. This feedback mode has only an indirect effect on the gas mass distribution (and hence on tracer particles), turning some fraction of the released thermal energy into kinetic energy and launching a quasar-like wind. 

In radio mode, a jet is launched from the AGN. \@The jet moves mass from the central cell only and spreads it into the jet and injects linear momentum, and energy. The released energy (and hence, momentum within the jet), as for the quasar mode, is proportional to the rest-mass accreted energy with an efficiency of $\varepsilon_{\rm f, R}$ , which is either taken as a constant value of 1 (to match the SMBH-to-galaxy mass relation~\citealp{dubois_self-regulated_2012}) or a varying function of the spin of the SMBH following the results of magnetically arrested discs (MADs) from~\cite{mckinney_general_2012} (see Dubois et al., in prep.\ for details). The jet is modelled by a ``capsule'' (a cylinder with spherical caps) of size $r_\AGN$, as illustrated in Fig.~\ref{fig:jet-scheme}. The radius of the jet $r_\AGN$ is usually set to a few times the cell resolution. The mass sent through the jet is proportional to the accreted mass onto the SMBH
\begin{equation}
  \dot{M}_\text{jet} = f_{\rm Load} \dot{M}_\SMBH,
\end{equation}
where $f_{\rm Load}$ is a mass-loading factor, usually \num{100}. The mass transported by the jet is distributed to all the cells intersecting with the capsule. Each cell $i$ receives a relative fraction $\psi_i$ of the mass (and of the injected linear momentum)
\begin{equation}
    \psi_i = \frac{\rho_i\int_\mathcal{I} e^{-r^2/2r_\AGN^2}\mathrm{d}^3 V}{\sum_j \rho_j \int_\mathcal{J} e^{-r^2/2r_\AGN^2}\mathrm{d}^3 V}\,,
    \label{eq:fraction-jet-psi}
\end{equation}
where $\mathcal{I}$ (resp. $\mathcal{J}$) is the volume of the intersection between the AGN capsule and the cell $i$ (resp. $j$) and $\rho_i$ is the cell mean density. The radius $r$ in Eq.~\eqref{eq:fraction-jet-psi} is the polar radius in the cylindrical frame centred on the AGN and aligned with its direction (it is the distance to the jet centre). This integral is computed approximately, using a numerical integration scheme. 

The tracer particles are interfaced with the jet model as follows. Each gas tracer particle in the cell $i$ containing the SMBH is moved into the jet volume with a probability of
\begin{equation}
  p_\text{jet} = \frac{\dot{M}_\text{jet}\Delta t}{M_i}.
  \label{eq:tracer-jet-proba}
\end{equation}
For each of these particles a random number $r$ is drawn from a uniform distribution {between 0 and 1}. If $r<p_\text{jet}$, the tracer is selected and moved into the jet.
The new position of the tracer $(x,y,z)$ is drawn randomly, $z$ being the coordinate in the direction of the jet; $x$ and $y$ are drawn from a normal distribution of variance $r_\AGN$ and $z$ is drawn uniformly between $-2r_\AGN$ and $2r_\AGN$. The algorithm uses a draw-and-reject method until one position inside the capsule is found. We note that the gas tracer distribution (as given by Eq.~\eqref{eq:tracer-jet-proba}) is consistent with the distribution of the gas sent through the jet (as given by Eq.~\eqref{eq:fraction-jet-psi})\footnote{In practice, the numerical evaluation of the integrals of Eq~\eqref{eq:fraction-jet-psi} may lead to small yet undetected discrepancies between the gas tracer and the gas distributions.}.

More details about the algorithm 
are given in Appendix~\ref{sec:algo}.

\section{Validations and tests}%
\label{sec:tests}
Let us now present various validation tests of the algorithm.  
Section~\ref{sec:idealized-tests} presents the results of idealised tests for gas-only tracer particles. Section~\ref{sec:astrophysical-tests} presents the results obtained from a cosmological zoom-in simulation of a galaxy with its SMBH at $z=2$ and provides the details of the observed distribution of tracer particles. Unless stated otherwise, the gas hydrodynamics is solved with an adiabatic index of $\gamma=5/3$ and the HLLC approximate Riemann solver~\citep{toro09}, applying the MinMod slope limiter on the linearly reconstructed states.

\subsection{Idealised tests}%
\label{sec:idealized-tests}

In this section, we introduce different idealised tests to confirm that the evolution of the gas is correctly tracked by gas tracers. Section~\ref{sec:diffusion} presents a simple two-dimensional (2D) advection of an overdensity to quantify diffusion effects. Sections~\ref{sssec:sedov-explosion} and \ref{sssec:KH-instab} present a Sedov-Taylor explosion and a Kelvin-Helmoltz instability and confirm that the gas tracers are able to accurately follow the motion of the gas for a strong shock case and a mixing layer of gas, respectively.
Section~\ref{sec:agn-feedback} presents an idealised halo with an AGN at its centre to confirm that the gas tracers correctly track the evolution of the gas in AGN jets.

\subsubsection{Uniform advection}%
\begin{figure}
    \centering
    \includegraphics[width=\columnwidth]{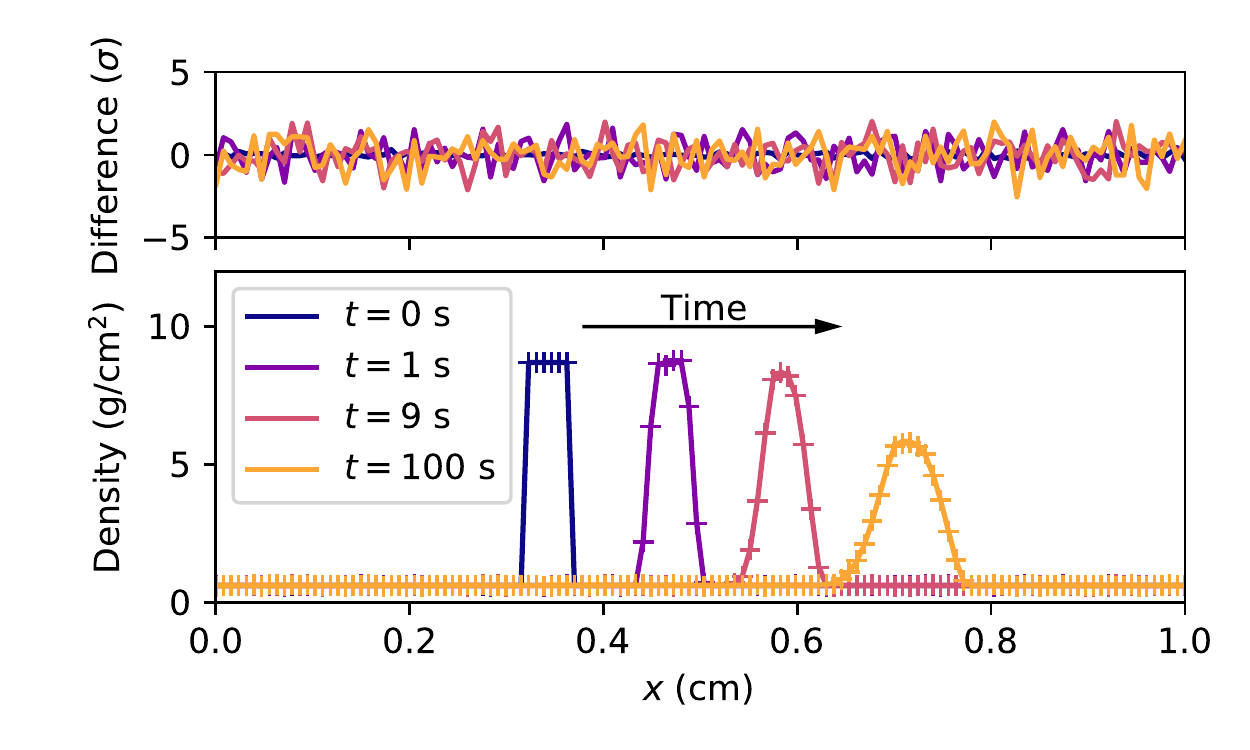}        \includegraphics[width=\columnwidth]{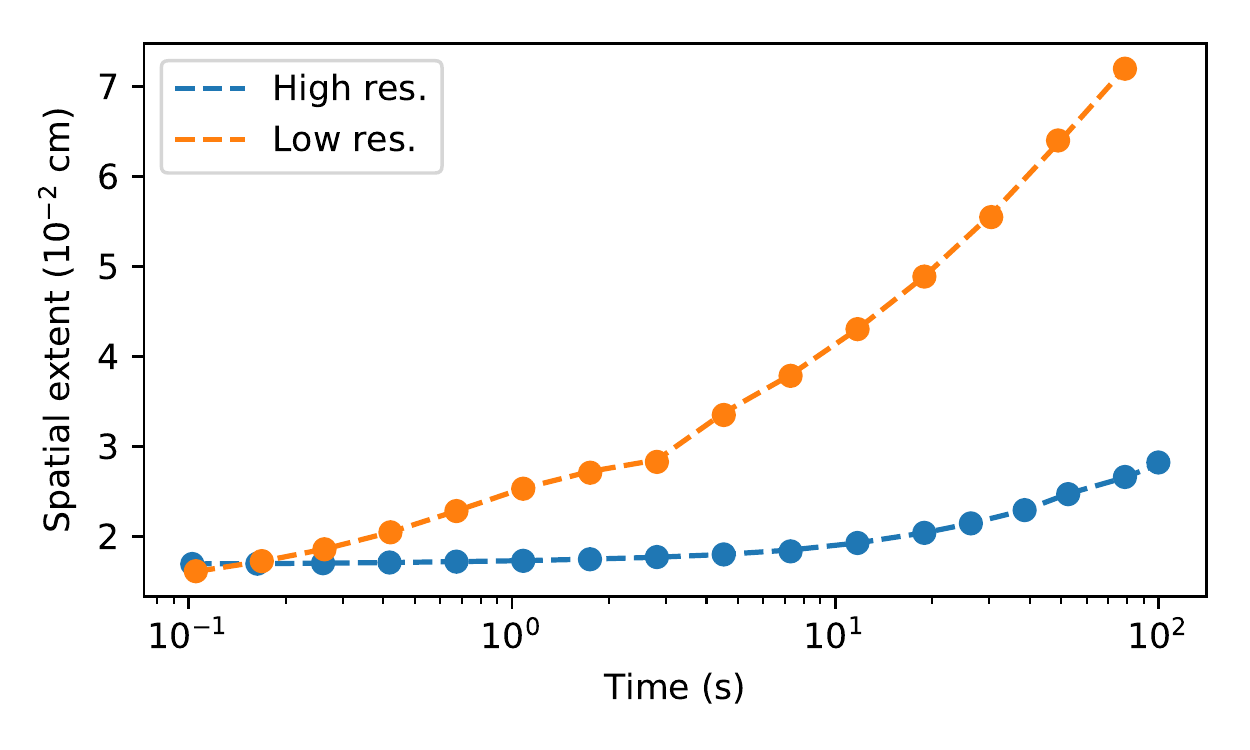}
    \caption{\textit{Centre:} Gas density profile (solid line) and gas density profile (plus symbols) at different times (reported in the legend). The profiles have been recentred and shifted horizontally by \SI{-0.12}{cm}, 0, \SI{0.12}{cm,} and \SI{0.24}{cm} for $t=0, 1, 9,$ and \SI{100}{s,} respectively. \textit{Top:} Relative difference between the gas and gas tracer density profiles in units of the expected noise level $\sigma=1/\sqrt{M_\cell/m_\tt}$. \textit{Bottom:} Evolution of the spatial extent of an advected overdensity as a function of time for the gas (dashed) and the gas tracer particles (dot symbols) for a high-resolution run (blue) and a low-resolution run (orange, see text for details). The difference shows no spatial dependence. The gas tracers diffuse exactly as the gas.}%
    \label{fig:simple_advection_profiles}
\end{figure}

\label{sec:diffusion}
In order to quantify the level of diffusion of MC tracers, we run a simulation similar to that run for Figure~6 of \cite{genel_following_2013}. The simulation is a region of \SI{1}{cm} in size with a constant density of \SI{1}{g/cm^3} and a velocity of \SI{0.01}{cm/s}. An overdensity of \SI{14}{g\per\cm^3} is set at $0<x<\SI{0.05}{cm}$. The sound speed is $c_{\rm s} = \SI{1.3}{cm/s}$ in the under-dense region and \SI{0.35}{cm/s} in the over-dense region. The simulation is performed on a uniform 2D $128^2$ grid including \num{250000} tracer particles, initially distributed in the same way as the gas. Due to the intrinsic numerical diffusion (advection error) of the hydrodynamical solver, the spatial extent of the overdensity increases as a function of time as it is advected away. This is illustrated in the central panel of Fig.~\ref{fig:simple_advection_profiles}. We note that the density profiles have each been shifted along their $x$ coordinate for visualisation purposes and do not reflect their real absolute position (in fact the rightmost peak travelled \SI{5}{cm} in \SI{100}{s}). The top panel of Fig.~\ref{fig:simple_advection_profiles} shows that, when rescaled by the expected noise level $\sigma \equiv 1/\sqrt{M_\cell/m_\tt} = 1/\sqrt{N}$ ($N$ is the expected number of tracer particles in the cell), the relative error between the gas tracers and the gas distributions shows no spatial modulation. Their distributions are the same with an extra factor that is entirely due to sampling noise, which in turn depends only on the local cell mass and the (constant) tracer mass.

In more quantitative terms, let us compare the time evolution of the spatial extent of the gas tracer overdensity to that of the gas. We rerun the simulation on a $32^2$ grid (low resolution) in addition to the previous run (high resolution). We compute the spatial extent by fitting a Gaussian function $\rho(x) = \rho_0 + H\exp(-(x-x_0)^2/(2\sigma_\rho^2))$ to the gas and gas tracer profiles, with free parameters $\rho_0$ the base density, $H$ the amplitude of the overdensity, $x_0$ the position of the overdensity, and $\sigma_\rho$ its spatial extent. The results are shown in the bottom panel of Fig.~\ref{fig:simple_advection_profiles}. As expected due to the numerical diffusion, the spatial extent of the overdensity increases as a function of time and the diffusion becomes larger when the resolution is decreased. In both cases, the Eulerian distribution of tracer particles is diffused exactly as much as the gas.\footnote{This result complements that of \cite{genel_following_2013}. Indeed we study here the diffusion of the Eulerian distribution of the tracer particles, while the original paper presents the Lagrangian diffusion of the tracer particles.}

\subsubsection{Sedov-Taylor explosion}%
\label{sssec:sedov-explosion}

We ran a classical Sedov-Taylor explosion in three dimensions and compare the gas density radial profile to the density profile of gas tracer particle. The simulation was performed on a coarse grid of $128^3$, refined on the relative variation of the density and of the pressure: a new level is triggered when the local relative variation of one of these quantities is larger than \SI{1}{\percent} with up to two levels of refinement. The simulation was initialised with a uniform density and pressure of \SI{1}{g\per cm^3} and \SI{e-5}{dyne\per cm^2} ,  respectively, and an over-pressure in the central cell of the box of \SI{6.7e6}{dyne\per cm^2}. \num{2900000} tracers, statistically uniformly distributed initially in the box, hence, with around $\sim 1.4$ tracer per initial cell.

\begin{figure}
  \centering
  \includegraphics[width=\columnwidth]{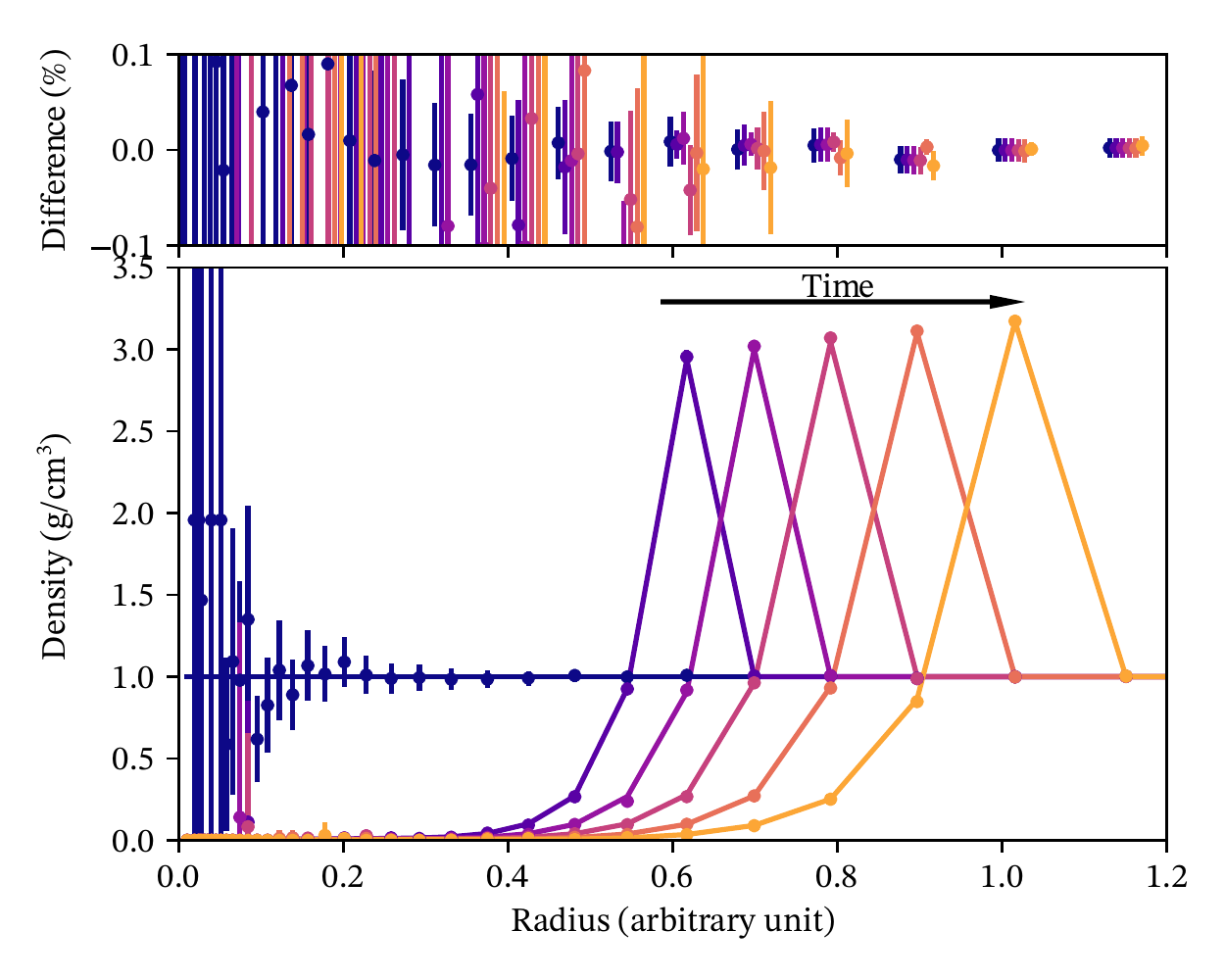}
  \caption{\textit{Bottom:} Radial profile at different times of a Sedov explosion (from blue to yellow) for the gas (solid lines) and the gas tracer (dots). The error bars are 2$\sigma$ errors. \textit{Top:} Relative difference between the gas profile and the gas tracer profile.  Data have been shifted by -0.25, -0.125, 0, 0.125 and 0.25 radius units respectively (from blue to yellow) so that one may easily distinguish the different data points. Details of the simulation are discussed in the text. The gas tracer particles are accurately advected with the gas.}%
  \label{fig:sedov}
\end{figure}

The evolution of the spherically averaged radial density profile of the gas and of the tracers is shown in Fig.~\ref{fig:sedov}. The tracer density has been computed by deposing the gas tracer mass in the nearest cell. The axes have been normalised so that the radius of the blast is one at the latest output. The error bars have been estimated assuming that the number of tracers per radial bin is given by a Poisson distribution. This assumption is discussed in more detail in Sect.~\ref{sec:gas-tracers}.

At all stages of the blast, the tracer particles radial profile matches that of the gas at percent levels. This is more easily seen in the top panel of Fig.~\ref{fig:sedov} where the relative difference between the gas tracer density and the gas density is plotted. The errors are all within a few percent and consistent with random fluctuations. As the explosion expands, the swept-up mass of gas in the shocked region increases. This is well tracked by the tracer distribution. Because the mass increases, the total number of tracer particles in the shock increases proportionally, causing the sample noise to decrease. In this particular test, the tracer distribution accurately reproduces that of the gas in the pre- (which is trivially that of the initial distribution) and post-shocked regions (shocked shell plus hot bubble interior). The noise level is a function of the number of tracer particles; its expected value is proportional to the total gas mass only.

The Sedov explosion is a reliable way of testing the ability of hydrodynamical codes to deal with shocks: more specifically it tests the ability of the code to capture the shock dynamics properly and also tests that the code resolves the shock interface with a few cells in a regime where the Mach number is largely above 1. Here, the gas tracer distribution has been shown to match that of the gas to a high degree of confidence, confirming that the gas tracers are correctly transported with the flow and are able to resolve shocks.

\subsubsection{Kelvin-Helmholtz instability}%
\label{sssec:KH-instab}

We ran a classical Kelvin-Helmoltz (KH) instability in three dimensions to compare the gas
density to the gas tracer density projected maps. The gas has an adiabatic index $\gamma=7/5$\footnote{This value is consistent with the adiabatic index of air at \SI{20}{\degree}.}.  The simulation is performed on a $128^3$ grid with a physical size of \SI{1}{cm} and a maximum level of refinement of $2^{10}$. Cells are refined based on the relative variation of the density: a new level is triggered when the local relative variation of the density is larger than \SI{1}{\percent}. Only hydrodynamics is included. The instability is initialised with two regions of left and right density of \SI{2}{g/cm^3} and \SI{1}{g/cm^3}, and of tangential velocity $u_{y,\rm L}=\SI{-1}{cm/s}$ (resp. $u_{y,\rm R}=\SI{1}{cm/s}$). The instability was initially triggered by adding a small damped sinusoidal perturbation of the perpendicular velocity field $u_{x}=u_0\cos\left(k (x-\lambda/2)\right) \exp(-k|x-x_0|)$, where $\lambda=\SI{0.25}{cm}$, $k=2\pi/\lambda$, $x_0=\SI{0.5}{cm}$ and $v_0=\SI{0.1}{cm\per s}$. Here \num{2900000} gas tracers were initially distributed in the box, so that their Eulerian distribution matched that of the gas.

\begin{figure}
    \centering
    \includegraphics[width=\columnwidth]{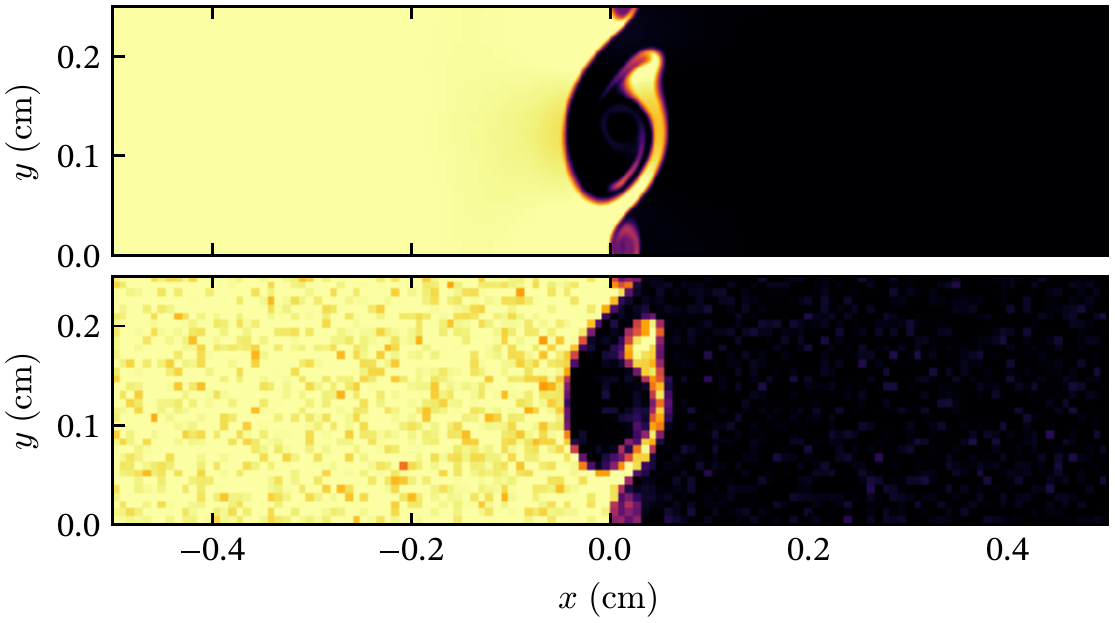}
    \caption{Projection of the density (\textit{top panel}) and  of the gas tracer density (\textit{bottom panel}) around a developing Kelvin-Helmoltz instability. To reduce the noise of the gas tracer projection, we have superposed the four projections of the forming rollers (each of size \SI{0.25}{cm}). The gas tracer distribution resembles that of the gas with extra noise due to their stochastic nature.}%
    \label{fig:KH_instab}
\end{figure}

Figure.~\ref{fig:KH_instab} shows a projection of the gas density and of the tracer density at time $t=\SI{0.3}{s}$, when the Kelvin-Helmoltz was already settled. The gas tracer distribution reproduces well the vortices found in the gas distribution, with extra noise due to the reduced number of tracer particles. 

\begin{figure}
    \centering
    \includegraphics[width=\columnwidth]{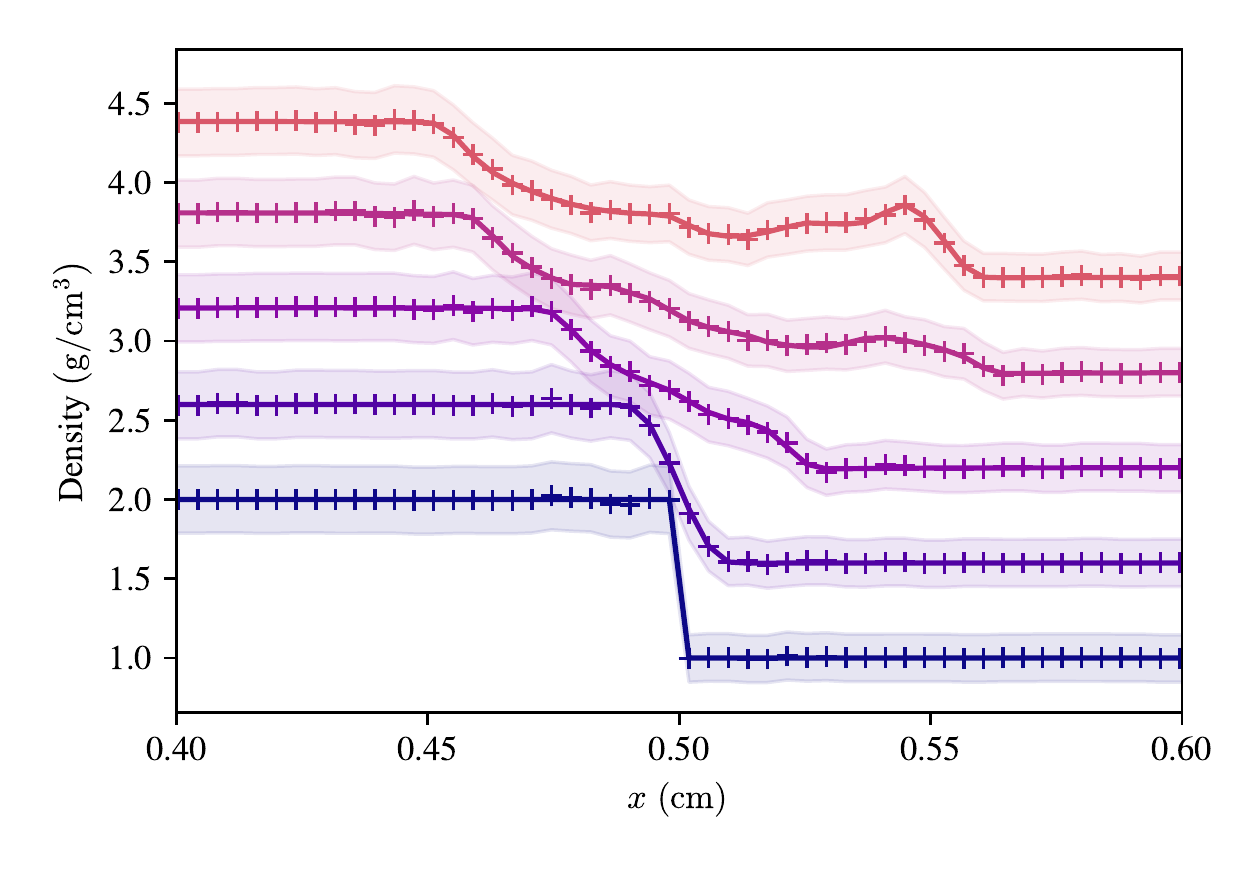}
    \caption{Evolution of the cross-section of the gas density (solid lines) and the gas tracer density (symbols and shaded regions) for the Kelvin-Helmoltz instability  at different times (from blue to red from the start to the end of the simulation at $t=\SI{0.3}{s}$). The profiles have been shifted vertically (each by \SI{0.6}{g/cm^3}) so that one may easily distinguish them from one another. The shaded regions are $\pm5 \sigma$, where $\sigma$ has been estimated using a Poisson sampling noise. The gas tracers are accurately following the diffusion of the gas.}%
    \label{fig:KH_instab_timeseries}
\end{figure}

The largest $k$ wave numbers of the perturbation are the first to grow following a KH growth timescale of $\tau_{\rm KH}=2\pi\mathcal{R}^{1/2}/(|\Delta u| k)$, with $\pm\mathcal{R}=\rho_{\rm R}/\rho_{\rm L}$, and $\Delta u=u_{y,\rm R}-u_{y,\rm L}$. 
Therefore, as time proceeds, larger rollers develop in the shear interface between the two phases of gas, and hence, the mixing layer spreads further.
We computed the evolution of the cross-section profile of the density at different times. The results are presented in Fig.~\ref{fig:KH_instab_timeseries}. The phase-mixing region grows as a function of time and the growth is correctly captured by the tracer particles that are able to track it within their intrinsic noise level.
Therefore, the gas tracer particles are able to correctly capture the KH shear instability leading to mixing of two gas phases.
Interestingly, the present algorithm does not lead to any relative diffusion between the gas and the tracers, as is illustrated quantitatively in Sect.~\ref{sec:diffusion}.

\subsubsection{AGN feedback}%
\label{sec:agn-feedback}

We subsequently tested the accuracy of the mass transfer for the jet mode of AGN feedback, which transfers part of the gas of the central cell to the surrounding cells within a ``capsule'' region (see Sect.~\ref{sec:agn-implem} for details).
We ran an idealised simulation of a halo with an AGN at its centre. The simulation is performed on a coarse grid of $128^3$, refined according to a quasi-Lagrangian refinement criterion: a cell is refined/derefined wherever the mass resolution is above/below \SI{1.4e7}{M_\odot} up to a maximum level of refinement of $12$. The box size is \SI{1.2}{Mpc}, hence with a minimum cell size of \SI{300}{pc}.
The max level of refinement is also enforced in all the cells closer than $4\Delta x$ from the SMBH, where $\Delta x$ is the minimum cell size.
The gas distribution follows a NFW~\citep{navarro_universal_1997} gas density profile, while the dark matter part follows a similar NFW profile modelled with a static gravitational profile (no back reaction of gas onto dark matter). 
 The NFW profile has parameters $V_{200}=\SI{200}{km/s}$ (at 200 times the critical density of a $H_0=\SI{70}{km\per s\per Mpc}$ Universe), a concentration of $c=6.8$, and is \SI{10}{\percent} gas. The gas is initially put at rest and at hydrostatic equilibrium. A SMBH of mass $M_{\SMBH,0}=\SI{3.5e10}{\Msun}$\footnote{We note that the SMBH mass is taken anomalously high for a typical halo mass of $M_{200}\simeq \SI{3e12}{M_\odot}$. This is chosen simply to get a sufficient power of the jet through the Bondi accretion rate given the NFW distribution of gas.} is set at the centre of the box and \num{e6} tracers are set in the cell containing the black hole. We force the AGN to be in jet mode with a fixed direction in space and boost its efficiency so that all the tracer particles are sent into the jet in one time step. The radius and height of the jet is $r_\AGN=\SI{50}{kpc}$. This value is much larger than usual values which are usually a few times the cell resolution (here typical values would be a few kiloparsecs). This is chosen so that the jet reaches cells at different levels of refinement and in other CPU domains. Within \SI{50}{kpc} of the AGN, there are \num{1200}, \num{24000}, \num{12000}, \num{13000} and \num{8000} cells at levels $2^8$ to $2^{12}$ ($\Delta x$ from \SI{5}{kpc} to \SI{0.3}{kpc}) so that the tracer particles are deposited in regions of different refinement level. This region also covers 8 of the 16 CPU domains used. This controlled test enables us to check that the distribution of tracers sent through the jet matches the expected distribution, in the presence of deep refinement and parallelism.

Let us first present the theoretical probability distribution function as a function of the distance to the jet and along the jet. We then compare theoretical figures to those of the simulation. The marginal probability density function (PDF) in the direction of the jet $r_\parallel$ is given by
\begin{equation}
  \label{eq:rparallel}
  p(r_\parallel) = \frac{1}{A}\begin{cases}
    \sqrt{e}-e^{r_\parallel^2/2r_\AGN^2} \,,& \text{if}\ |r_\parallel|<r_\AGN\,, \\
    \sqrt{e}-1\,, & \text{if}\ r_\AGN < |r_\parallel| < 2r_\AGN,
  \end{cases}
\end{equation}
where
\begin{equation}
  A = 2\sqrt{e}r_\AGN\left(2+\sqrt{2}F(1/\sqrt{2})-1/\sqrt{e}\right).
\end{equation}
Here $F$ is Dawson's integral. The marginal PDF in the radial direction $r_\perp$ is
\begin{equation}\label{eq:rperp}
  p(r_\perp) = \frac{r_\perp e^{-r_\perp^2/2r_\AGN^2}\left(1+\sqrt{1-r_\perp^2/r_\AGN^2}\right)}
  {r_\AGN^2 \left(2-\sqrt{2}F(1/\sqrt{2})-1/\sqrt{e}\right)}.
\end{equation}
The marginal PDF in the radial distribution is similar to a $\chi$ distribution with two degrees of freedom with an extra factor due to the two spherical caps: more particles are found close to the centre of the jet since the capsule is more extended close to its centre.

\begin{figure}
  \centering
  \includegraphics[width=\columnwidth]{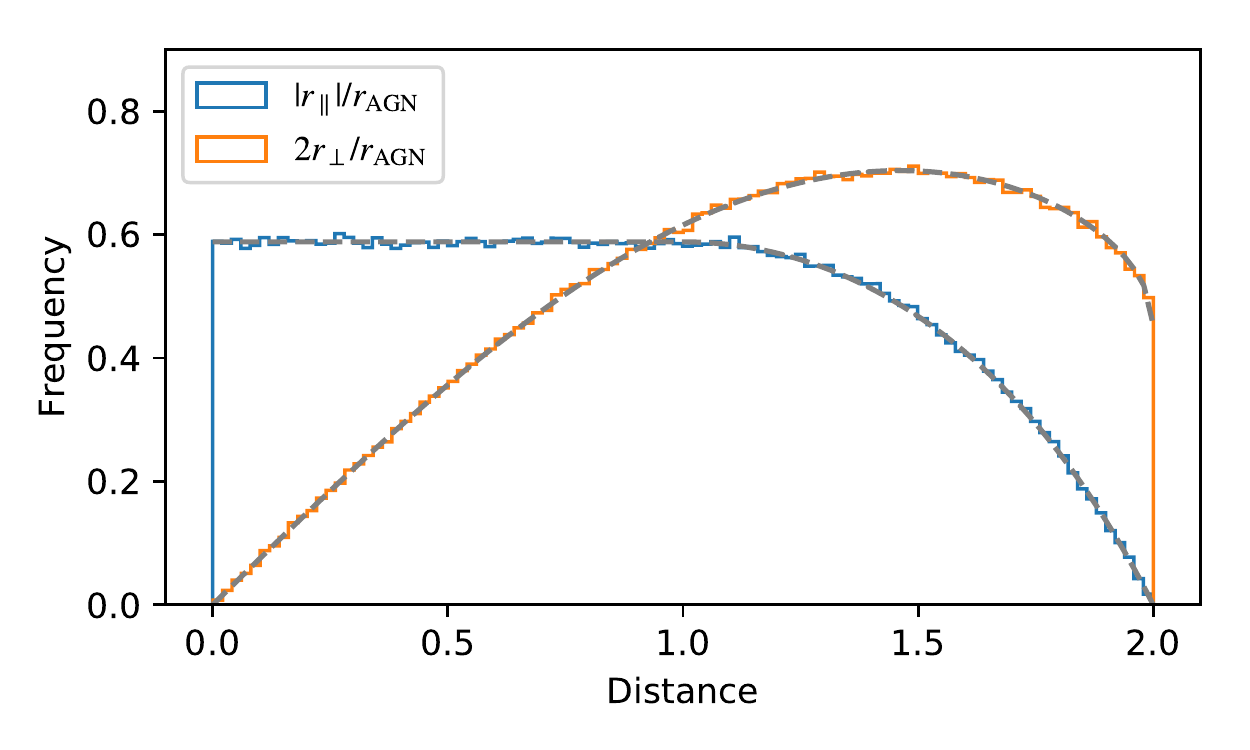}
  \caption{Distribution of particles moved by a jet before any hydrodynamical time step has occurred. Shown is the parallel distribution marginalised over the plane of the jet (blue) and the radial distribution marginalised over the direction of the jet (orange) vs.\ the expected theoretical distributions from Eqs.~\eqref{eq:rparallel} and \eqref{eq:rperp} (dashed grey).
  The abscissa is in units of $r_\AGN$ in the parallel direction and in units of $r_\AGN/2$ in the radial direction. The distribution of gas tracers sent into the jet perfectly matches the expected one.}%
  \label{fig:jet-tracers}
\end{figure}

Figure~\ref{fig:jet-tracers} presents the results from the comparison of the simulation to the expected distribution. The distribution in the radial direction has been rescaled by a factor of two to span the same range as in the parallel direction. Theoretical curves (Eqs.~\eqref{eq:rparallel} and \eqref{eq:rperp}) are in very good agreement with the observed distributions, confirming that the algorithm is distributing tracer particles correctly in jets. In addition we have also run the same idealised simulation without forcing the AGN efficiency. We report that the tracer mass flux is equal to the gas mass flux. This confirms that the physical model of the jet is accurately sampled by the tracer particles interacting with it, both in terms of its mass and for its spatial distribution.

\begin{figure*}
\includegraphics[width=\textwidth]{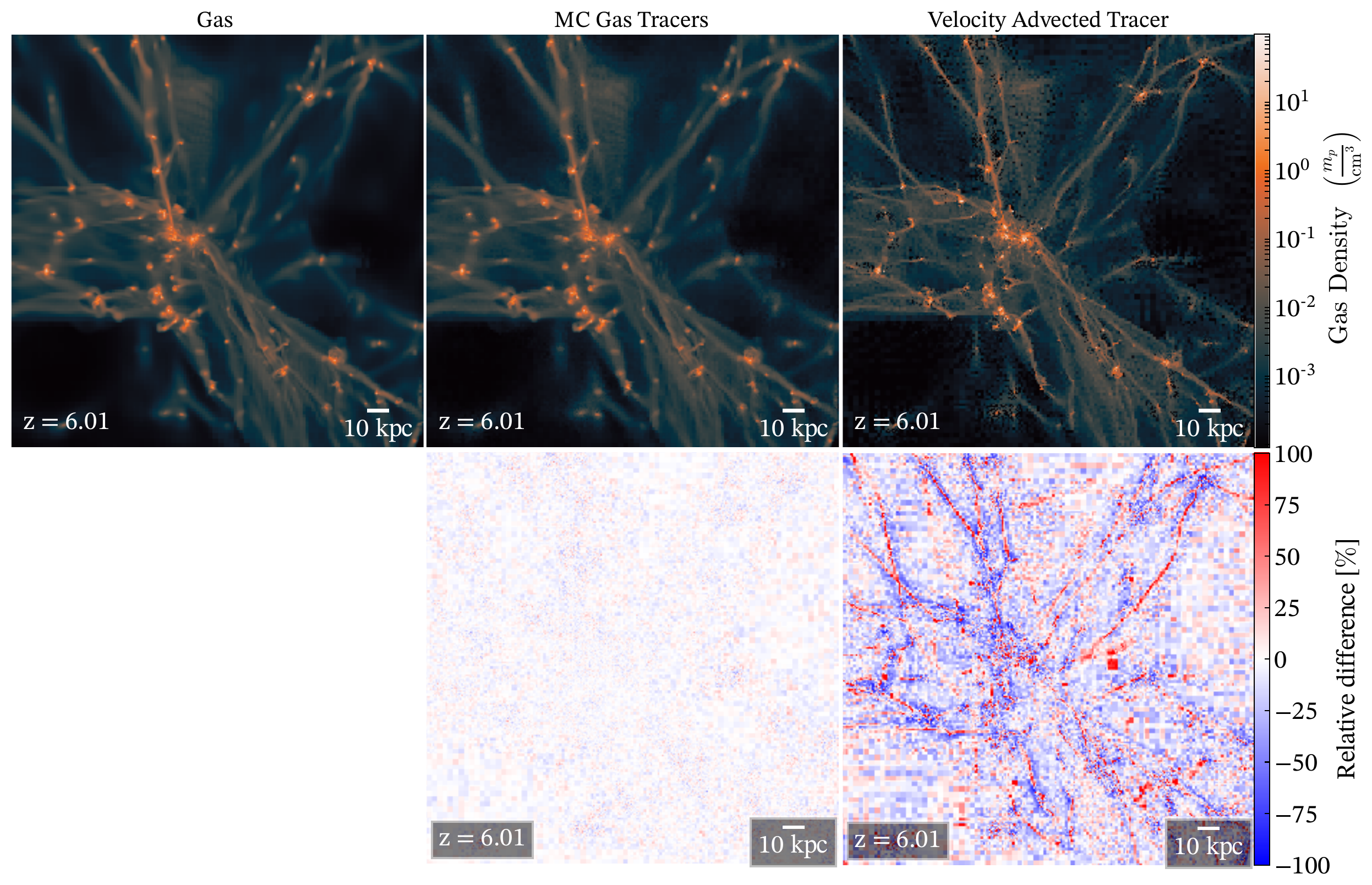}
  \caption{\textit{Top:} Density weighted projection of the gas density in a cosmological simulation (\emph{left}), of the  velocity tracer distribution (\emph{right}), and of the MC gas tracer distribution (\emph{centre}). All the plots share the same colour map. \textit{Bottom:} The relative difference between the tracer and the gas. Velocity tracers accumulate in convergent regions (e.g.\ filaments, nodes). The MC gas tracer distribution reproduces more accurately that of the gas than velocity tracers.}%
  \label{fig:vel-vs-gas}
\end{figure*}

\subsection{Astrophysical test}%
\label{sec:astrophysical-tests}

We have run a \SI{50}{cMpc/h}-wide cosmological simulation down to $z=2$ zoomed on a group of mass \SI{1e13}{\Msun} at $z=0$, where the size of the zoom in the Lagrangian volume of initial conditions is chosen to encapsulate a volume of two times the virial radius of the halo at $z=0$.
We start with a coarse grid of $128^3$ (level 7) and several nested grids with increasing levels of refinement up to level 11.
The adopted cosmology has a total matter density of $\Omega_{\rm m}=0.3089$, a dark energy density of $\Omega_{\Lambda}=0.6911$, a baryonic mass density of $\Omega_{\rm b}=0.0486$, a Hubble constant of $H_0=\SI{67.74}{km\per s\per Mpc}$, a variance at \SI{8}{Mpc} $\sigma_8=0.8159$, and a non-linear power spectrum index of $n_{\rm s}=0.9667$, compatible with a Planck 2015 cosmology \citep{planck_collaboration_planck_2015}.

The simulation includes a metal-dependant tabulated gas-cooling function following~\cite{sutherland_cooling_1993} allowing the gas to cool down to $T\sim\SI{e4}{K}$ via Bremsstrahlung radiation (effective until $T \sim \SI{e6}{K}$), and via collisional and ionisation excitation followed by recombination (dominant for $\SI{e4}{K} \leq T \leq \SI{e6}{K}$). The metallicity of the gas in the simulation is initialised to
$Z_0=\SI{e-3}{Z_\odot}$ to allow further cooling below $\SI{e4}{K}$ down to $T_{\rm min}=\SI{10}{K}$.
Reionisation occurs at $z=8.5$ using the~\cite{haardt_radiative_1996} model and gas self-shielding above \SI{e-2}{m_p\per\cm^3}. Star formation is allowed above a gas number density of $n_0= \SI{10}{H\per cm^3}$ according to the Schmidt law and with an efficiency $\varepsilon_{\rm ff}$ that depends on the gravo-turbulent properties of the gas \cite[for details, see][]{kimm_feedback_2017,trebitsch_fluctuating_2018}. The main distinction of this turbulent star-formation recipe with the traditional star formation in {\sc Ramses} \citep{rasera_history_2006} is that the efficiency can approach and even exceed \SI{100}{\percent} (with $\varepsilon_{\rm ff} > 1$ meaning that stars are formed faster than in a free-fall time). The stellar population is sampled with a \cite{kroupa_variation_2001} initial mass function, where $\eta_{\rm SN}=0.317$ and the yield (in terms of mass fraction released into metals) is $0.05$. The stellar feedback model is the mechanical feedback model of \cite{kimm_towards_2015} with a boost in momentum due to early UV pre-heating of the gas following~\cite{geen_detailed_2015}. The simulation also tracks the formation of SMBHs and the evolution of AGN feedback in jet mode (radio mode) and thermal mode (quasar mode) using the model of \cite{dubois_self-regulated_2012}. 
The jet is modelled in a self-consistent way by following the angular momentum of the accreted material and the spin of the black hole~\citep{dubois_black2_2014}. The radiative efficiency and spin-up rate of the SMBH is then computed using the MAD results of \cite{mckinney_general_2012}.

We have a minimum roughly constant physical resolution of \SI{35}{pc} (one additional maximum level of refinement at expansion factor $0.1$, $0.2$, and $0.4$), a star particle mass resolution of $m_{\star,\rm res}=\SI{1.1e4}{\Msun}$, a dark matter (DM) particle mass resolution of $m_{\rm DM, res}=\SI{1.5e6}{\Msun}$, and gas mass resolution of \SI{2.2e5}{\Msun} in the refined region.
A cell is refined according to a quasi-Lagrangian criterion: if $\rho_{\rm DM}+\rho_{\rm b}/f_{\rm b/DM}>8 m_{\rm DM,res}/\Delta x^3$, where $\rho_{\rm DM}$ and $\rho_{\rm b}$ are respectively the DM and baryon density (including stars plus gas plus SMBHs), and where $f_{\rm b/DM}$ is the cosmic mean baryon-to-DM mass ratio. The max level of refinement is also enforced in all cells closer than $4\Delta x$ from any SMBH, where $\Delta x$ is the minimum cell size.
We add tracer particles in the refined region with a fixed mass of $m_\tt = \SI{2.0e4}{\Msun}$ ($N_\text{tot}\approx\num{1.3e8}$ particles). There is on average 0.55 tracers per star and 22 per initial cell. Cells of size \SI{35}{pc} and density \SI{20}{cm^{-3}} contain on average one tracer per cell.

\subsubsection{Velocity tracers versus\ Monte Carlo tracers}%
\label{sec:vel-vs-MC}

In addition to the above simulation, we ran the exact same one replacing each MC tracer with a velocity-advected tracer. This simulation was performed down to $z=6$ and compared to the fiducial one. Both have a similar gas distribution, confirming that the tracer particles are indeed passive.\footnote{They have however an indirect impact on stochastic processes such as star formation and SN feedback as they impact the random number generator (hence the outcome of these random processes will vary depending on how many and where the tracer particles are).} At this redshift, \SI{99}{\percent} of the baryons are still in the gas phase (\SI{0.72}{\percent} in stars and \SI{8e-5}{\percent} in SMBHs), meaning that the comparison between MC tracers (that can be transferred into stars) and velocity tracers is fair when looking at cosmological scales. Since the velocity tracers have not been linked to star formation or SMBHs, we expect significant discrepancies within galaxies, where the gas-to-star ratio is much smaller.

The top panels of Fig.~\ref{fig:vel-vs-gas} show projections of the density-weighted density of gas (top left panel), of MC tracers (top-centre panel), and of velocity-advected tracers (top-right panel). The distribution of the MC tracers resembles that of the gas with extra noise due to sampling noise. All the prominent structures in the gas are also present in the MC tracer distribution. On the other hand, the velocity tracer distribution is much sharper than that of the gas. The velocity tracers aggregate in converging flows (filaments and centres of galaxies) while MC tracers do not (they aggregate in high-mass regions, as expected). At such large scales, the origin of the discrepancy is an intrinsic issue of velocity tracers. This test shows that on a qualitative level, the MC tracers have a distribution that is in much better agreement with the gas distribution than the velocity advected tracers. The relative difference between the gas distribution and the tracer distribution is presented in the bottom panels of Fig.~\ref{fig:vel-vs-gas}. The relative difference between the MC tracer density and the gas density (bottom central panel) is significantly smaller than the relative difference between the velocity advected tracer density and the gas density (bottom right panel). The latter is also much more biased: the velocity advected tracer density in convergent flows (e.g.\ filaments) can be up to an order of magnitude larger than the gas density, while in the vicinity of converging regions, the velocity advected tracer density is largely underestimated (e.g.\ around filaments). On the contrary, the MC tracer density is found to be in better agreement with the gas density and is not biased.

\subsubsection{Gas tracers}%
\label{sec:gas-tracers}

\begin{figure*}
  \includegraphics[width=\textwidth]{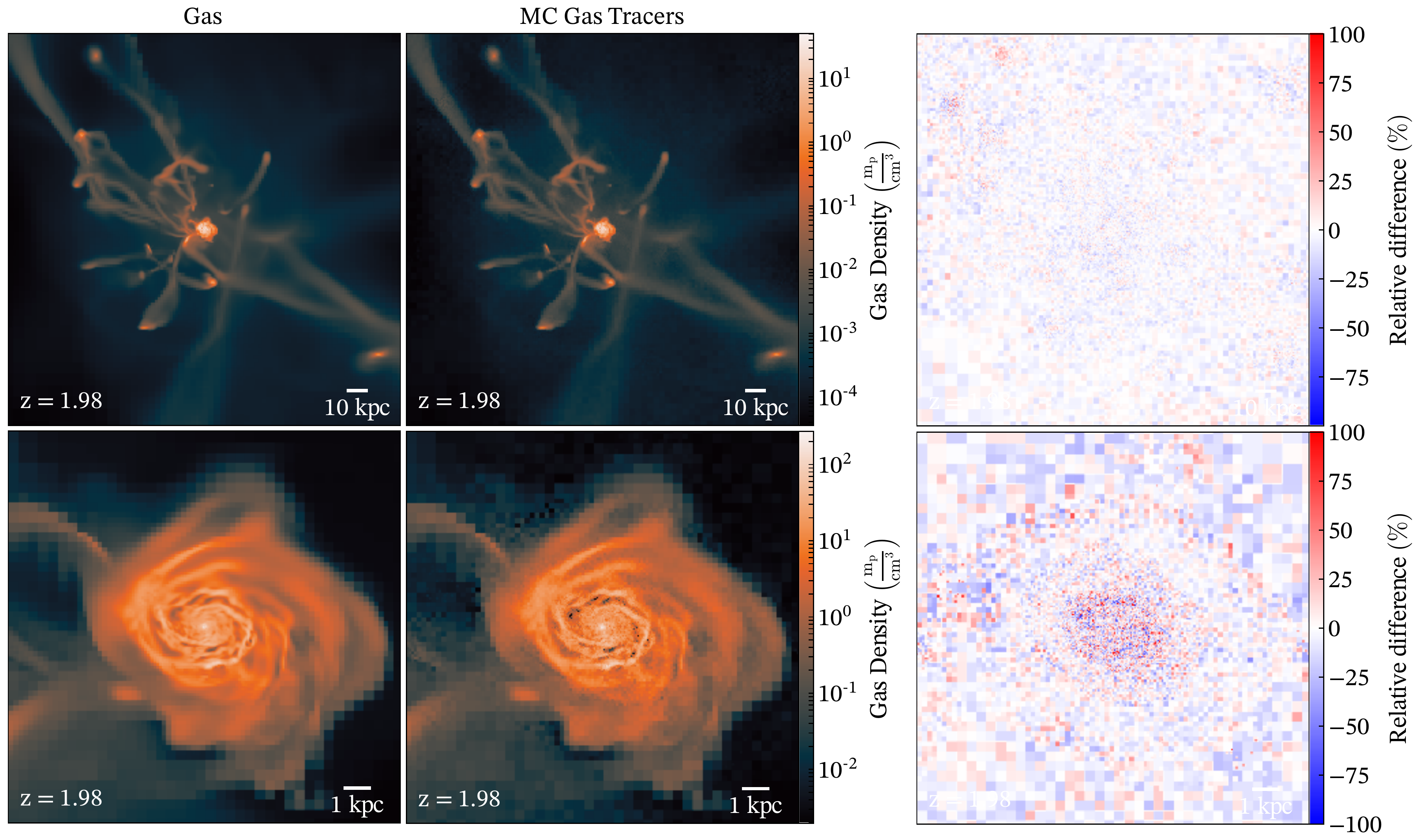}
  \caption{Density-weighted projection of the gas density (\textit{left}), of the gas tracer
    density (\textit{centre}), and of their relative difference (\textit{right}) along the $x$ axis around the most massive galaxy of the cosmological
    simulation at $z=2$. \textit{Top:} The large-scale structure of the gas; data have been selected within \SI{200}{kpc} of the centre. \textit{Bottom:} Zoom on the central galaxy; data have been selected within \SI{10}{kpc} of the centre of the galaxy. The MC tracer density is similar to that of the gas. The radial modulations are due to differences in cell mass at fixed cell resolution: massive cells (closer to the centre at fixed resolution) are best sampled by the MC tracers.}%
  \label{fig:galaxy-xyz}
\end{figure*}

\begin{figure}
  \centering
\includegraphics[width=\columnwidth]{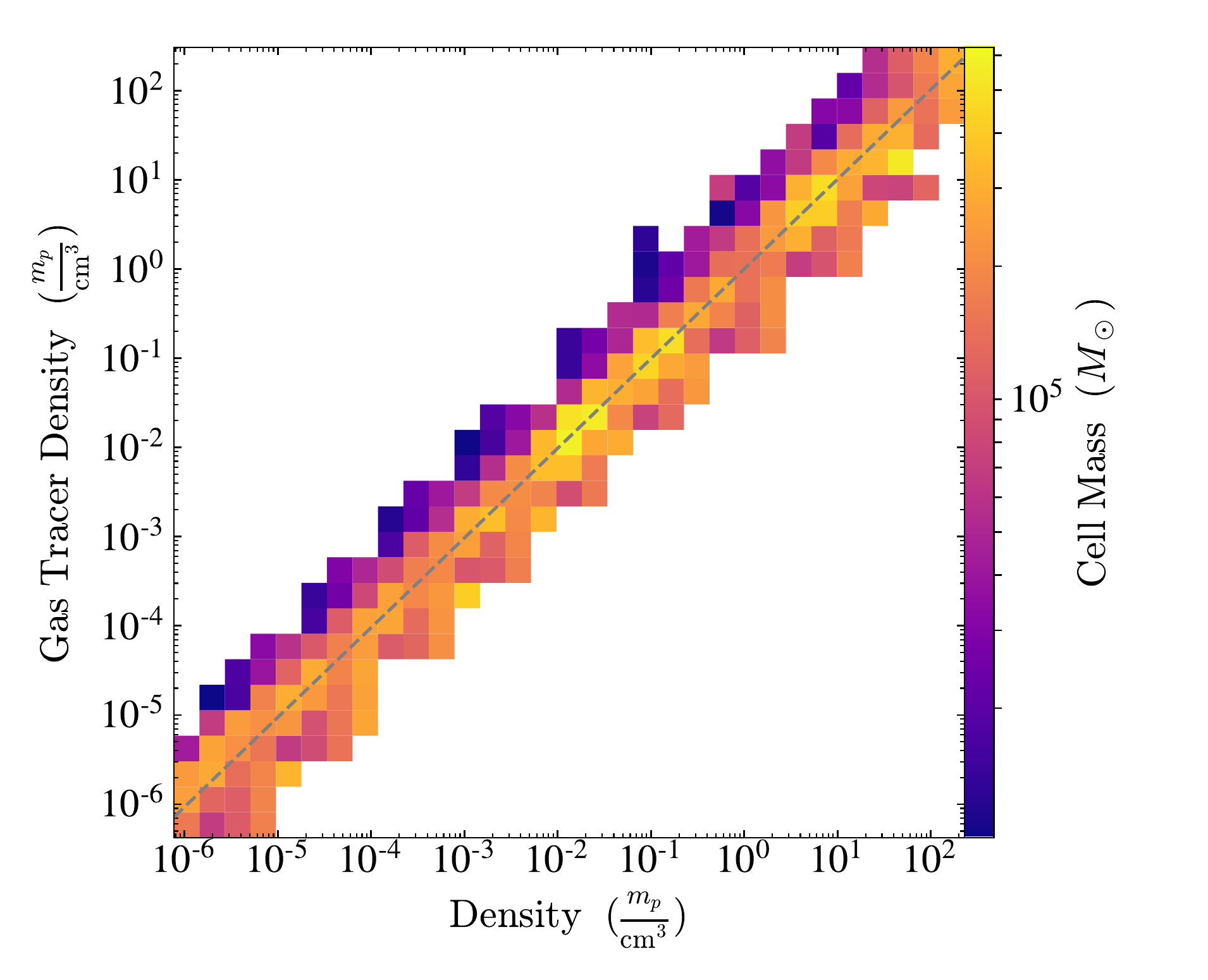}
  \caption{Gas density  vs.\ gas tracer density, colour coded by cell mass. The grey dashed line shows the one-to-one relation. The gas and gas tracer densities match on nine orders of magnitude.}%
  \label{fig:rho-rhotracer}
\end{figure}

As we have seen, velocity tracer particles are a less reliable tracer of the actual gas density compared to MC tracer particles, and this can already be seen on cosmological scales. Therefore, we now continue to explore only the distribution of MC tracer particles with respect to the actual distribution of baryons.
Figure~\ref{fig:galaxy-xyz} shows the density-weighted projected gas density and cloud-in-cell interpolated gas tracers around the zoomed galaxy of the simulation. Visual inspection reveals that the gas tracer distribution matches that of the gas with additional noise. All structures with a contrast above the noise level are reproduced by the gas tracers.
More quantitatively, Fig.~\ref{fig:rho-rhotracer} shows the density of tracers versus the density of gas for the entire available range of gas densities (i.e. 9 orders of magnitude); the expected one-to-one relation is seen, with some scatter due to MC sampling noise.

\begin{figure}
  \centering
  \includegraphics[width=\columnwidth]{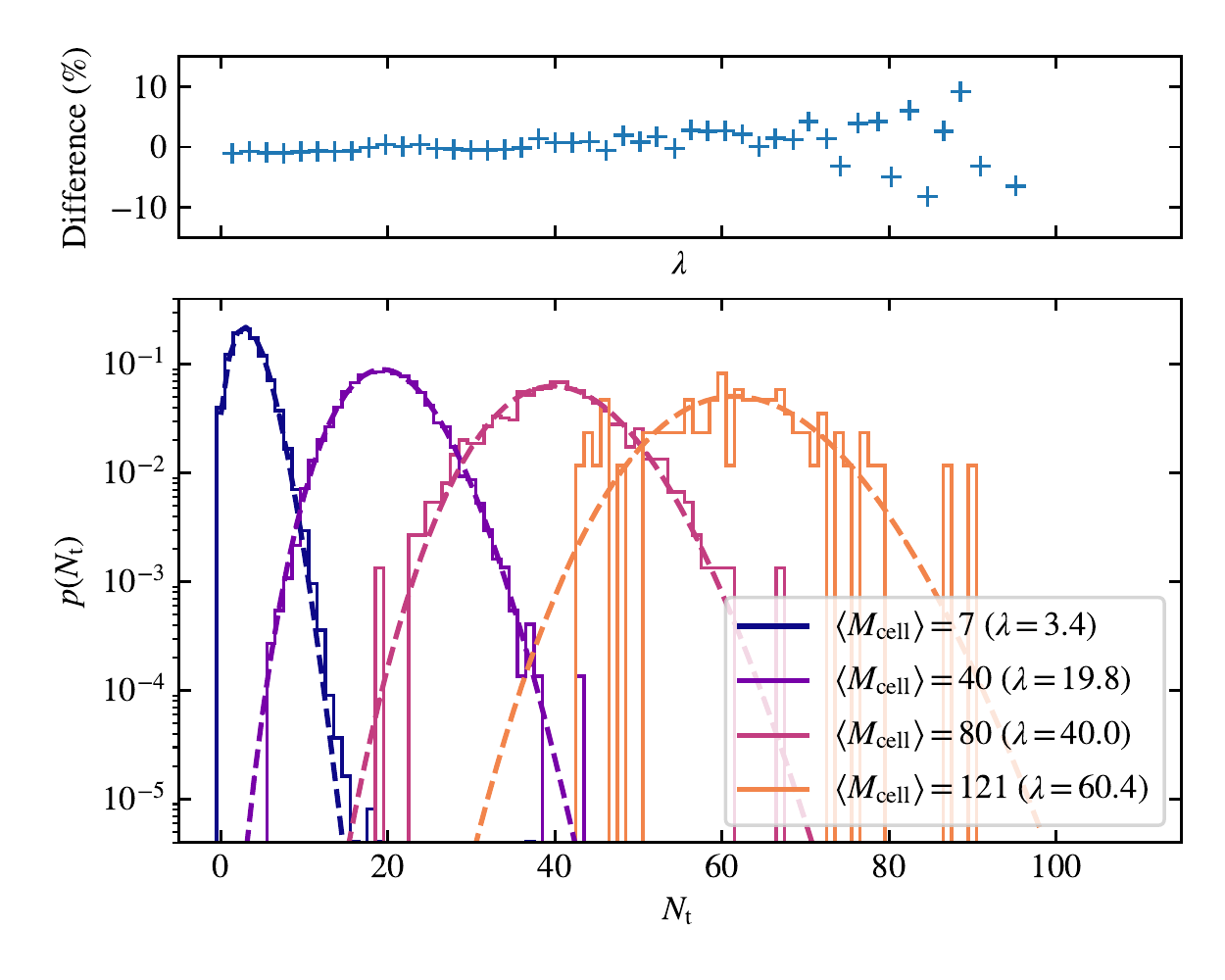}
  \caption{\textit{Bottom:} Distribution of the number of gas tracers for different cell-mass bins as observed in the simulation (solid lines) vs.\ a Poisson distribution with parameter $\lambda = \langle M_\cell\rangle/m_\tt$ (dashed lines, reported in the legend). \textit{Top:} Relative difference between the observed mean number of tracer particles and the expected one, $\lambda,$ as a function of $\lambda$. For all cells, the distribution of the number of gas tracers per cell is given by a Poisson distribution with parameter $\lambda$.}%
  \label{fig:gas-tracer-distribution}
\end{figure}

More quantitative results can be obtained by computing the statistical properties of the gas tracer population. A cell of mass $M_\cell$ is expected to contain on average $M_\cell/m_\tt$ tracers. For a sample of cells of similar masses, we expect the mean number of tracers per cell to be $\lambda \equiv \langle M_\cell \rangle / m_\tt$. The distribution of the number of tracers per cell in the simulation is shown in Fig.~\ref{fig:gas-tracer-distribution} for different cell-mass bins. Within a cell-mass bin, the number of tracers $N_\tt$ can be seen to be very well approximated by a Poisson distribution with parameter $\lambda$
\begin{equation}
  p_\lambda(N_\tt=k) = \frac{\lambda^{k} e^{-\lambda}}{k!}.
\end{equation}
To confirm this observation, we compared the mean number of tracers per cell to the expected number $\lambda$ in the top panel of  Fig.~\ref{fig:gas-tracer-distribution}. For all cell masses, the mean number of tracer particles per cell is accurately described by its expected Poisson distribution. At large values of gas mass within a cell (right of the plot), the scatter in the histogram count is due to the small number of massive cells in the simulation. Indeed, these cells can only be found in the most refined regions (otherwise they would be refined into smaller cells) where they also tend to be converted into stars.

In the following we assume that the gas tracer distribution is given by a Poisson distribution with parameter $\lambda=\langle M_\cell\rangle/m_\tt$. This yields a simple rule of thumb to estimate the precision of the tracer scheme. The accuracy of the Eulerian distribution of the tracer can be written $1/\sqrt{\lambda}\sim\sqrt{m_t/M_\mathrm{cell}}$.


\subsubsection{Star formation and feedback}

\begin{figure*}
  \includegraphics[width=\textwidth]{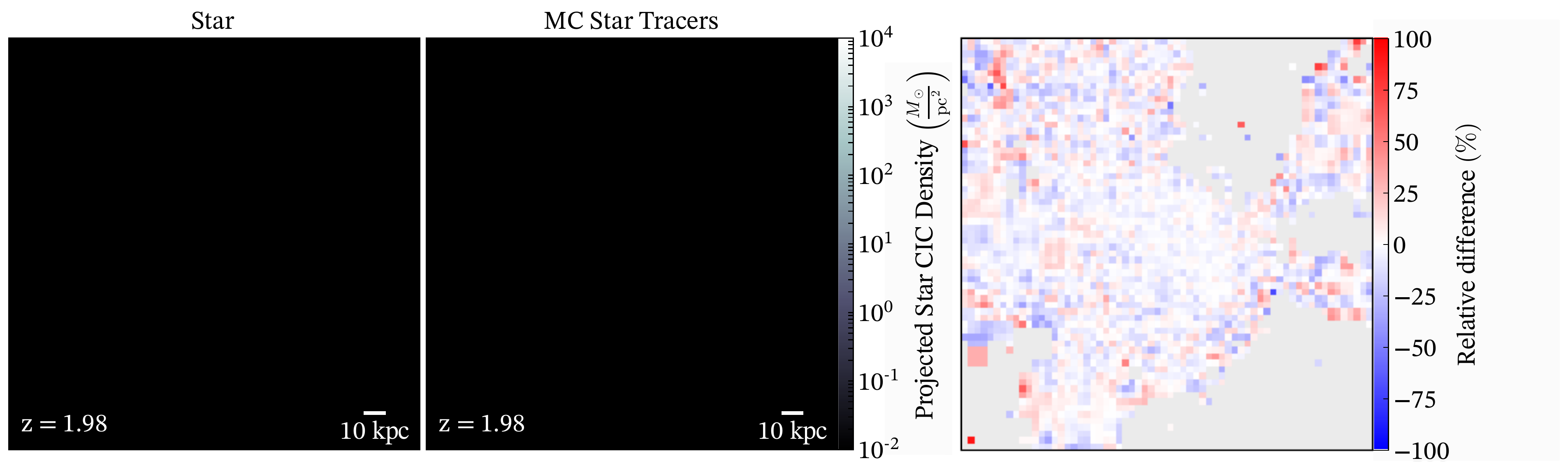}
  \caption{Stellar surface density (\textit{left}), star-tracer surface density (\textit{centre}), and relative difference (\textit{right}). The data are the same as in Fig.~\ref{fig:galaxy-xyz}. In the difference map, regions where no stars are found are indicated in grey. The star and star-tracer distributions are in very good agreement; their difference shows no spatial dependence. The noise level is higher than in Fig.~\ref{fig:galaxy-xyz} at large radii where the star surface density is smaller than the gas surface density, hence the star mass distribution is less resolved than the gas.}%
  \label{fig:galaxy-xyz-star}
\end{figure*}

\begin{figure}
  \centering
 \includegraphics[width=\columnwidth]{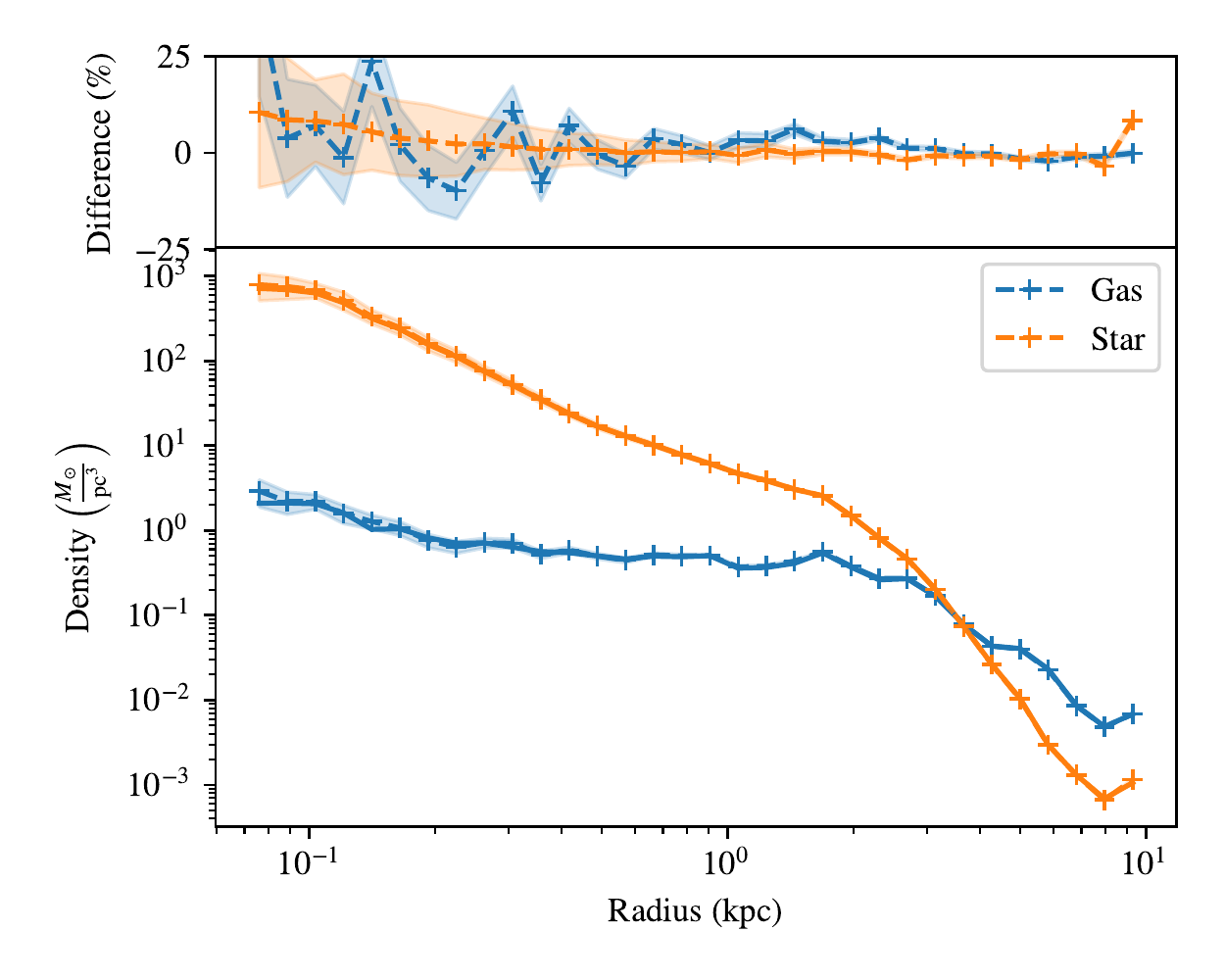}
  \caption{\textit{Bottom:} Radial profile of the gas density (solid blue) and star density (solid orange)  vs. the gas tracer density (blue cross) and the star-tracer density (orange cross). The error bars are given by a Poisson sampling noise. \textit{Top:} Relative difference between the baryon and the tracer profiles. The tracers match their baryon counterpart at a few percent level.}%
  \label{fig:radial-profile-galaxy-star}
\end{figure}

Figure~\ref{fig:galaxy-xyz-star} shows the integrated stellar mass and star-tracer mass around the zoomed galaxy of the cosmological simulation. Both distributions are visually in agreement and feature the same spatial distribution. At large radii where the star density is smaller than the gas density ($r\gtrsim \SI{4}{kpc}$, see Fig.~\ref{fig:radial-profile-galaxy-star}), the noise level of the star-tracer distribution is larger than that of the gas. This is due to the fact that small masses are poorly resolved by the MC tracers. Close to the galactic centre, the increasing star density induces a larger star-tracer density, and therefore, at fixed resolution, a smaller noise sampling. This is illustrated by the right panel of Fig.~\ref{fig:galaxy-xyz-star}, where the centre of the plot shows smaller fluctuations than at large radii. More quantitative results are presented below.

We first present the analytical distribution of tracer particles for stars and for the number of tracers released in SN events, derived from first principles.
When a star particle is formed, each tracer in the cell containing the newly created star particle is attached to the star particle and has a probability of $p_\star\equiv M_{\star,0} / M_\cell$  of becoming a ``star tracer'', where $M_{\star,0}$ is the mass of the newly created star particle.\footnote{We note that in practice the star particles have a mass that is a multiple of the stellar mass resolution.} Because $M_{\star,0} < M_\cell$ -- a star particle cannot be formed with more material than what is available -- this probability is well defined: $0 < p_\star < 1$. When the heavy stars in a star particle go into SN, they yield $\eta M_\star$, and the mass of the corresponding star particle becomes $M_\star=(1-\eta)M_{\star,0}$. The star tracers are then returned to the gas with a probability of $\eta$.
Before the SNe explode, the distribution of tracers for an individual star particle is given by a binomial distribution with parameters $N_i$ (the initial number of tracer in the cell where the star particle formed) and $p_\star$
\begin{equation}
  p_\text{form}(N_i;N_f=k) = \binom{N_i}{k}p_\star^k(1-p_\star)^{N_i-k}.
\end{equation}
The number of tracer particles released in the SN event reads
\begin{equation}
  p_\text{SN}(N_f; N=k) = \binom{N_f}{k} \eta^k(1-\eta)^{N_f-k},
\end{equation}
where $N_f$ is the number of star tracers in the star particle before the SN explosion. The number of tracers in the star particle after the SN has exploded is, thus, given by a binomial distribution of parameters $N_i$ and $(1-\eta)p_\star,$
\begin{equation}
  \label{eq:binom-star-tracer}
  p_\star^\mathrm{f}(N_i;N=k) =  \binom{N_i}{k} \left((1-\eta)p_\star\right)^k\left(1-(1-\eta)p_\star\right)^{N_i-k}.
\end{equation}

In the limit where the $N_i$ becomes large and $(1-\eta)p_\star$ small, Eq.~\eqref{eq:binom-star-tracer} converges mathematically to a Poisson distribution with parameter $N_i(1-\eta)p_\star$.

\begin{figure}
  \centering
  \includegraphics[width=\columnwidth]{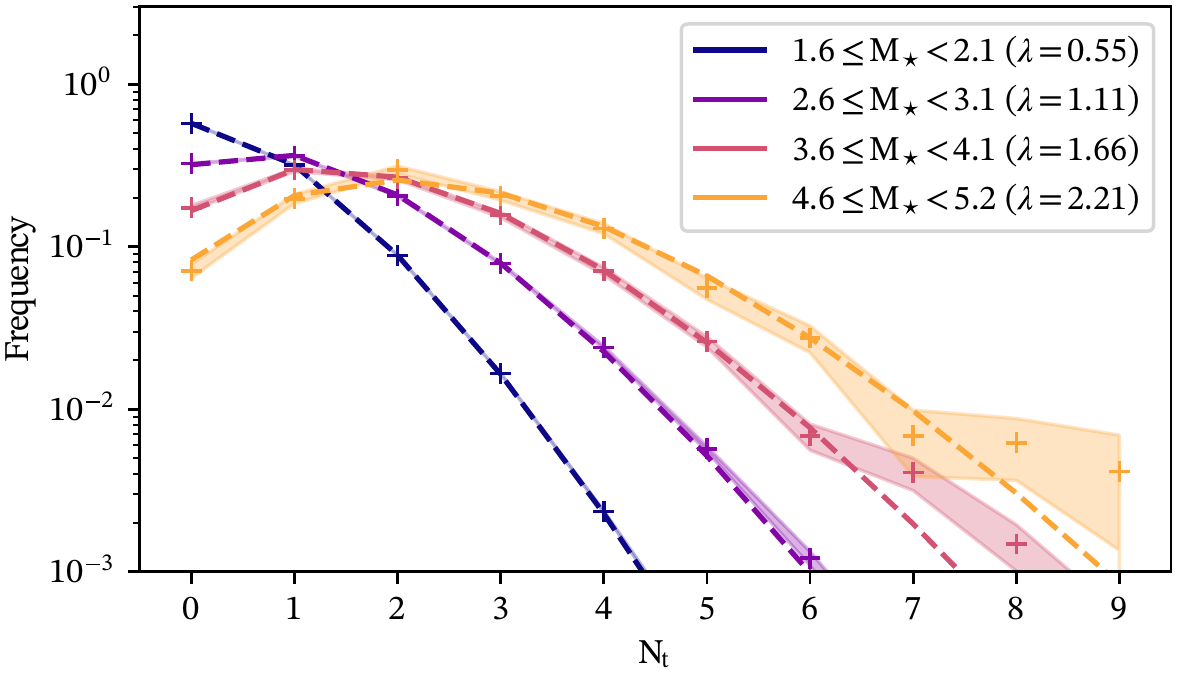}
  \caption{Distribution of the number of star tracers per star for different star particle mass bins (in units of \SI{e4}{\Msun}) as observed in the simulation (symbols and shaded surfaces) vs.\ as given by a Poisson distribution with parameter $\lambda = \langle M_\star\rangle /m_\tt$ (dashed). The error bars have been estimated using a bootstrap method. For all stars, the distribution of the number of star tracers per star is approximated by a Poisson distribution with parameter $\lambda$.}%
  \label{fig:star-tracers}
\end{figure}

Now, we compare the expected distribution of tracer particles to the measured one. Figure~\ref{fig:star-tracers} presents the distribution of the number of tracer particles per star particle for different star particle mass bins. The number of star tracers per star particle can be seen to be well approximated by a Poisson distribution with parameter $\lambda = \langle M_\star\rangle /m_\tt$. There is a clear deviation at the tail of the distribution which displays  an excess of probability. This is however expected as when a star forms in a cell, a significant part of the cell mass is converted into the star, so that $p_\star\approx 1$. Because usually $(1-\eta)\approx 0.9$, the product $p_\star(1-\eta)$ is also of order unity. At the same time, cells where stars form have a typical mass of $\SI{e4}{\Msun}\sim m_t$, meaning that they contain only {a few} gas tracers at star formation. Therefore, we expect  a  significant deviation from a Poisson distribution, as the requirement for Eq. (\ref{eq:binom-star-tracer}) to converge to a Poisson distribution is not met. This argument is reinforced by the fact that, compared to light stars (e.g.\ the blue curve of Fig. 15), the most massive stars  have a more top-heavy distribution (e.g.\ the red curve) than a Poisson distribution. Indeed, these  massive stars are relatively more massive than their parent cell, meaning that the parameter $p_\star$ is larger. In the simulation, star formation is only activated for cells above a given (fixed) density threshold. This is usually achieved at the maximum resolution, causing cells experiencing star formation to have typically the same mass, and therefore the same number of gas tracer particles, regardless of the mass of the forming stars. Consequently, the massive star particle  distribution is indeed  less Poissonian than that of the light stars, since their $p_\star$ is larger at fixed $N_i$. Figure~\ref{fig:star-tracers} is in qualitative agreement with this.

\subsubsection{SMBH evolution}

Using our cosmological simulations, we have checked that the total mass of SMBH tracer particles ($M_\mathrm{t\ SMBH,tot}=(\num{3.5}\pm\num{0.3})\times\SI{e6}{M_\odot}$\footnote{The uncertainty has been estimated using a 1-$\sigma$ Poissonian noise.}) matches that of SMBH in the simulation ($M_\mathrm{SMBH,tot}/(1-\varepsilon_r)=\SI{3.1e6}{M_\odot}$) at the 10
\% level, up to an $\varepsilon_r$ factor. This factor is due to the mass lost by the accreted material as it falls onto the black hole. This mass is radiated away and lost to the simulation. Because the tracer particles have a fixed mass in our implementation, they are unable to capture the mass energy that is radiated. However, one could store the value of $\varepsilon_r$ at accretion time onto each tracer to be able to reconstruct the exact mass that the SMBH tracer represents.

\subsection{Bi-modal accretion at high redshift: a science case for tracer particles}%
\label{sec:science_case}

\begin{figure}
    \centering
    \includegraphics[width=\columnwidth]{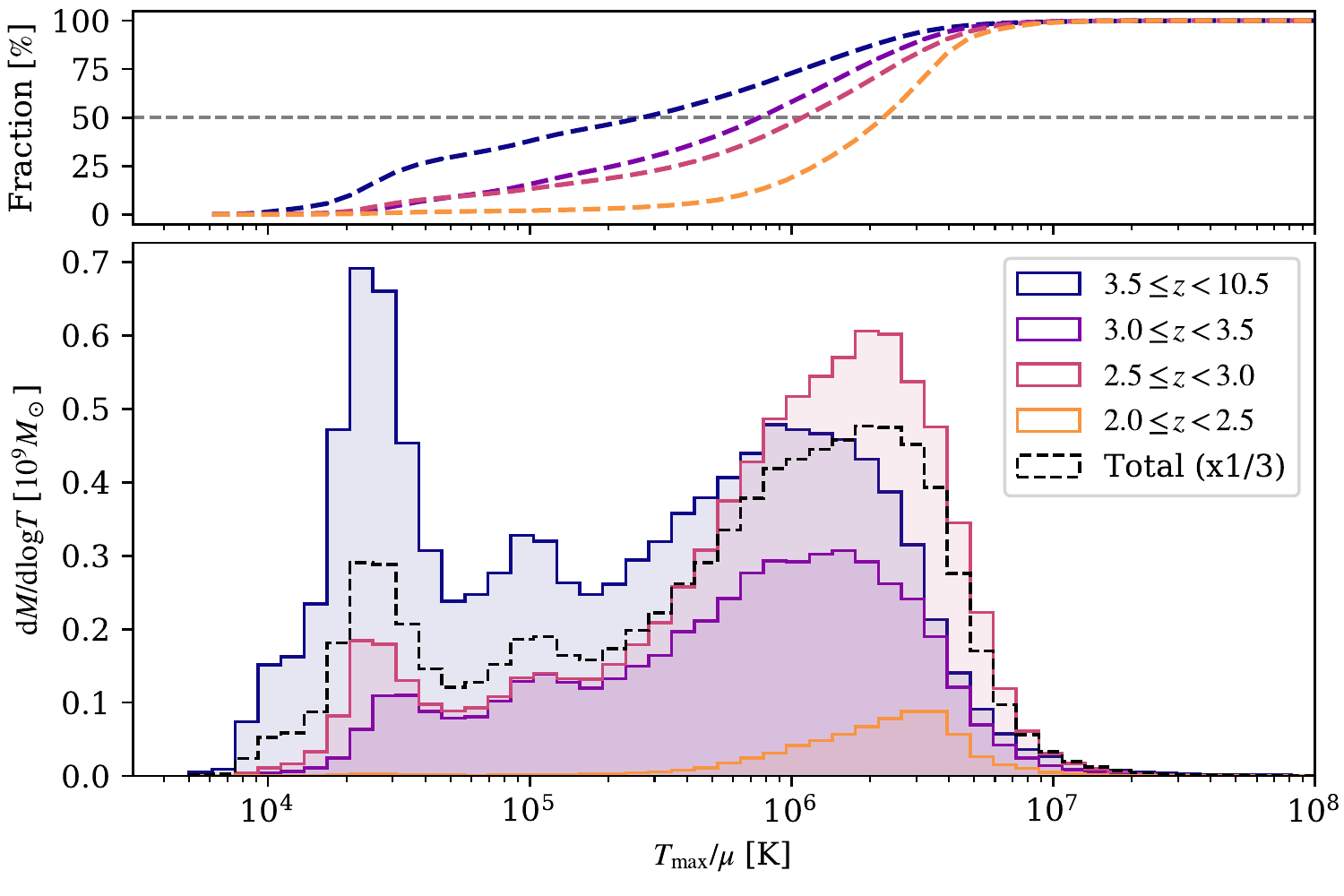}
    \caption{(\textit{Bottom:}) Histogram of the maximum temperature of the gas accreted onto the central galaxy between different redshifts (from early accretion time in blue to late accretion time in yellow). (\textit{Top:}) Cumulative distribution of the gas temperature.
    Only the gas-forming stars within the virial radius are selected. The total distribution integrated over the total accretion time is shown with the black dashed line in the bottom panel. The total distribution has been rescaled by a factor of one third for visualisation. The halo has two modes of accretion: a cold and a hot mode. At high $z$ the cold mode dominates and at low $z$ the hot mode dominates.}%
    \label{fig:max_temperature_histogram_different_times}
\end{figure}

Low-mass galaxies (embedded in halos $M_{\rm h}\lesssim 10^{11}\, \rm M_\odot$) exhibit a significant amount of ``cold-mode'' cosmological accretion made of cold gas streaming in narrow filaments with a temperature typically below $T_\mathrm{max}\lessapprox\SI{e5}{K}$~\citep{birnboim_virial_2003,keres_how_2005,ocvirk_bimodal_2008,nelson_moving_2013,nelson_accretion_2016}.
A ``hot-mode'' phase made of gas that was shock heated before entering the virial radius ($T_\mathrm{max}\sim \SI{e6}{K}$) appears in halos with higher mass. At early times ($z > 2.5$), the accretion is dominated by the cold mode. As time goes by, halos grow in mass so that an increasing fraction of the gas heats up before entering the halo. The outcome of this is a decrease of the relative importance of cold accretion compared to  hot accretion. By $z\lessapprox 2$, most of the accreted material comes from the diffuse hot phase.
Hence, getting access to the Lagrangian history of the stars and of the star-forming gas is key to pinning down the origin of gas acquisition in galaxies.

We revisit this result using {\sc ramses} and the MC tracer particles. Using the cosmological simulation of Sect.~\ref{sec:astrophysical-tests}, we study the accretion of gas as a function of time around the central galaxy. We select all the gas tracers that end up in star particles (not the star-forming gas) at $z=2$ and $r<0.1 R_\mathrm{vir}$. The halos were detected using the AdaptaHOP halo finder \citep{aubert_origin_2004-1}. For the positioning of the centre of the DM halo, we start from the first AdaptaHOP guess of the centre (densest particle in the halo) and from a sphere the size of the virial radius of the halo; we use a shrinking sphere \citep{power_inner_2003-1} by recursively finding the centre of mass of the DM within a sphere \SI{10}{\percent} smaller than the previous iteration. We stop the search once the sphere has a size smaller than $\simeq \SI{100}{pc}$ and take the densest particle in the final region. Twenty neighbours are used to compute the local density. Only structures with a density greater than 80 times the average total matter density and with more than 200 particles are taken into account. The original AdaptaHOP finder is applied to the stellar distribution in order to identify galaxies with more than 200 particles. Their Lagrangian history is reconstructed in post-processing from the 132 equally spaced ($\Delta t = \SI{25}{Myr}$) outputs, and the thermodynamical properties of the gas are extracted from the local gas cell value. For each tracer particle, the maximum temperature $T_\mathrm{max}$ reached before falling into the virial radius is recorded. The infall time is defined as the last inward crossing of the virial radius. The merger tree is computed following~\cite{tweed_building_2009-1}. The procedure only selects tracer particles falling onto the galaxy in the gas phase. This excludes gas tracers tracking gas that formed stars in satellite galaxies but includes gas from wet mergers. Figure~\ref{fig:max_temperature_histogram_different_times} presents the temperature distribution of the accreted gas for different bins of infall time. At early times (blue lines, $z\gtrsim 3$) the accretion is bi-modal. About \SI{50}{\percent} of the gas is accreted via the cold mode, as shown in the top panel of Fig.~\ref{fig:max_temperature_histogram_different_times}. At later redshifts ($z\lesssim 2.5$), the accretion becomes dominated by the hot mode. The relative importance of the cold accretion decreases and the distribution become less and less bimodal, until it is eventually entirely dominated by the hot mode. This is in qualitative agreement with the findings of \cite{keres_how_2005} though the exact quantitative amount of cold versus hot accreted gas relies significantly on i) the numerical scheme to model gas dynamics \cite{nelson_moving_2013} and ii) the modelled feedback processes \citep{dubois_blowing_2013}.

Caution should be taken here: contrary to what was done in the original study, only  the accretion onto a single galaxy is investigated. In particular, our results are sensitive to the particular accretion and merger history of that galaxy, which impact the temperature distribution of the gas. In order to achieve a fairer comparison, one would have to run a full cosmological simulation and study the gas accretion of the full population within the box. While this would now technically be possible thanks to the new tracer algorithm, it is nonetheless well beyond the scope of this paper.

\section{Performance}%
\begin{figure}
    \centering
    \includegraphics[width=\columnwidth]{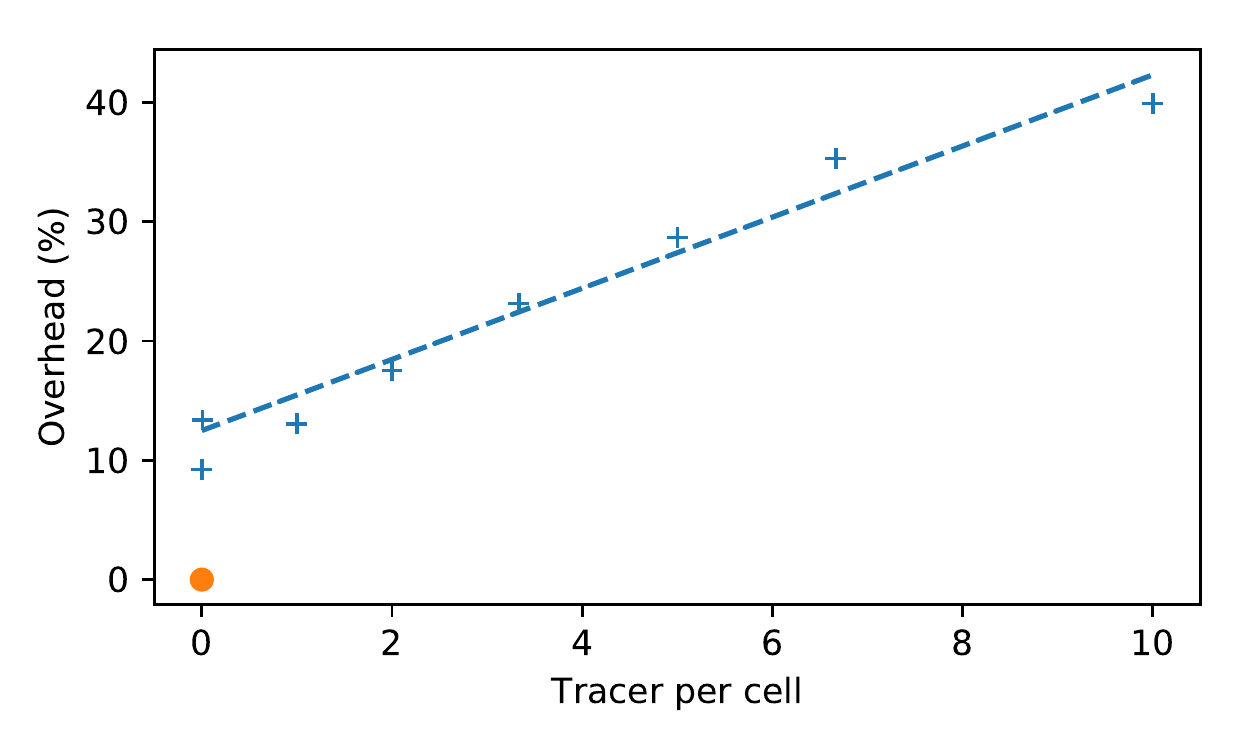}
    \caption{Overhead as a function of the number of tracer particles per initial cell (symbols). The orange symbol is the simulation with the tracer deactivated. The data (excluding the run with the tracer deactivated) have been fitted with a linear function (dashed line). The estimated overhead (slope of the fit) is $\sim\SI{3}{\percent}$ per tracer per initial cell with an extra constant of $\sim\SI{10}{\percent}$.}%
    \label{fig:benchmark}
\end{figure}

\label{sec:performance}
To quantify the performance of the tracer particles and their associated CPU overhead (defined as the excess of computation time required by the tracer particles), we restarted the simulation of Sect.~\ref{sec:astrophysical-tests} at redshift $z=2$, while varying the numbers of tracer particles to test the scaling of the algorithm. At restart, we decimate the tracer population to keep only 67, 50, 33, 20, 10, or \SI{0.1}{\percent} of the initial population (in the gas, star, and black holes). We also run a simulation with no tracer but all the tracer routines activated (\verb|t0|) and a simulation with no tracer and the tracer routines deactivated (\verb|notracer|). The parameters of the runs are presented in the first three columns of Table~\ref{tab:benchmark}. The run time is defined as the total run time divided by the number of steps. The overhead is defined as the relative increase of the run time with respect to the run \verb|not|. All the runs were stopped after two iterations of the coarse time step (about $\sim\SI{2000}{s}$ of run time, $\sim \SI{2.8}{Myr}$ of simulation time). The results are also plotted in Fig.~\ref{fig:benchmark}.

By comparing the two runs \texttt{t0} and \texttt{notr}, we conclude that the tracer particle machinery adds a constant cost of about 10\% to the computation. This is due to the fact that the tracer particles require the fluxes at the interface of each cell (six quantities per cell) to be stored, which then have to be communicated between CPUs. In addition, there are multiple loops that iterate over all the cells and all the particles (see Sect.~\ref{sec:implementation} for more details). In principle, this could be optimised by setting tracer particles in their own linked list, but we exploited the particle machinery available in {\sc Ramses}, and treated tracer particles just like standard particles (star or DM) with respect to code structure. In the following, the computation overhead will be expressed in terms of the number of tracer per initial cell: $N_\tt/N_{\mathrm{cell},i}$, where $N_\tt$ is the number of tracer particles and $N_{\mathrm{cell}, i}$ is the number of initial (gas) cells. 

The runs with tracers show that the total run time starts increasing with the number of tracer particles per cell\footnote{We note that here the number of cells is the one in the refined regions, not the initial number of cells.} when this number becomes of the order of $\sim 0.1$ tracer per initial cell. Above this threshold, the run time scales roughly linearly with the number of tracer per initial cell. We have run the simulation on the Occigen supercomputer with 672 cores (28 nodes of 24 cores). Each node is made of two Intel Haswell 12-Core E5-2690 V3s\footnote{See \href{http://ark.intel.com/products/81713/Intel-Xeon-Processor- E5-2690-v3-30M-Cache-2_60-GHz}{Intel-Xeon-Processor- E5-2690}.} running at a clock frequency of \SI{2.6}{GHz}. The nodes are wired together with a DDR Infiniband network (\SI{20}{Gbit\per\s}). The code was compiled with the Intel Fortran compiler version 17.0 and OpenMPI 2.0.2. In this setup the overhead is \SI{3}{\percent} per tracer per initial cell. For example the run \texttt{t100} with 10 tracer per initial cell had a \SI{40}{\percent} overhead. Part of the overhead is due to the tracer particles themselves (moving, generating random numbers, etc.). Another part is due to the load balancing. Indeed, in this simulation, tracer particles are only found in the zoomed region, which is already the most CPU-intensive region. Our simulation can be seen as a worst-case scenario for the tracer particles.
In general, let us write the conservative formula giving an estimate of the overhead induced by the tracer particles
\begin{equation}
    \frac{\Delta t}{t} = 0.03 \left(\frac{N_\tt}{N_{\cell,i}}\right) + 0.1,
\end{equation}
where $t$ is the run time and $\Delta t$ the extra cost induced by the tracer particles. Here, $N_\tt$ and $N_{\cell,i}$ are the total number of tracer particles and the total number of initial cells, respectively.

\begin{table}
    \centering
    \begin{tabular}{lS[table-format = 8.0]S[table-format = 1.3]|S[table-format = 4.0]S[table-format = 2.1]}
        {Name} & {Absolute} & {Tracer}  & {Run time} & {Overhead} \\
        {}       &{number}  & {per cell}& {(s)}      & {(\%)}\\
        \midrule
        \verb|t100| &  129325116 &            10    &     1310 &     39.9 \\
         \verb|t67| &   86214303 &             6.7  &     1270 &     35.3 \\
         \verb|t20| &   64656206 &             5    &     1210 &     28.7 \\
         \verb|t33| &   43104621 &             3.3  &     1160 &     23.1 \\
         \verb|t20| &   25861310 &             2    &     1100 &     17.5 \\
         \verb|t10| &   12929077 &             1    &     1060 &     13.1 \\
        \verb|t0.1| &     130250 &             0.01 &     1060 &     13.4 \\
          \verb|t0| &          0 &             0    &     1020 &      9.2 \\
         \verb|not| &          0 &             0    &      940 &      {-}
    \end{tabular}
    \caption{Run time per coarse time step for the different runs. The run \texttt{notr} was performed with no tracer particles {and with} all the tracer particle routines deactivated. The column ``Tracer per cell'' is the number of tracer particles per initial cell in the zoomed region. The ``Overhead'' column contains the run-time overhead defined with respect to the \texttt{notr} run.}%
    \label{tab:benchmark}
\end{table}

\section{Conclusions}%
\label{sec:conclusions}

We present a new implementation of tracer particles in the {\sc Ramses} AMR code based on the Monte Carlo approach from~\cite{genel_following_2013}. It has been interfaced with the most common physical models used in cosmological simulations (star formation and stellar feedback, SMBH growth and AGN feedback). We have shown that the Lagrangian history of the gas is accurately reconstructed by testing the accuracy of the tracer distribution in an advection-dominated problem and in a diffusion-dominated problem. The gas tracer distribution  matches that of the gas, even in complex situations that involve subgrid models.
We have also provided a comparison of the new MC tracer particles to the previous velocity-based implementation and showed that the new version largely outperforms the accuracy of the previous one. We have made a detailed study of the distribution of tracer particles in a zoom-in cosmological simulation including state-of-the art subgrid model physics (cooling, star formation, SN feedback, SMBHs, and AGN feedback) and show that:
(i) in each cell, the gas tracer distribution is given by a Poisson distribution with parameter $\lambda = M_\cell / m_\tt$; and
(ii) for each star, the number of star tracers can be approximated by a Poisson distribution with parameter $\lambda = M_\star / m_\tt$.
The properties of the Poisson distribution give a simple rule to estimate the sampling noise of the tracer particle, as the noise can be represented by $1/\sqrt{\lambda}$. In turn this should allow users to quantify how many particles are needed to reach their sought accuracy. 
We have also shown that the gas tracer particles sample exactly the intrinsic numerical diffusion of the Godunov solver.
To highlight the assets of tracer particles in a realistic setting, they were implemented in the problem of cold flow  accretion at high redshift. The known bi-modality in the temperature of gas was recovered.  

The performance of the algorithm was explored. In a zoom-in full physics cosmological simulation, the run time grows roughly linearly with the number of tracer particles per cell. The overall impact on computation time is estimated to be $\sim\!\!3\ \%$ per tracer per initial cell plus a constant computation time overhead of $10\%$, regardless of the number of tracer particles. These figures should serve as upper limits on the computation time. The performance of the scheme could be optimised by using two separate linked lists for the tracer particles and the other particles, as is done in {\sc arepo} \citep{genel_following_2013}.
Implementing these possible improvements will be the subject of future studies. Presently, the performance is significantly lower than that reported in the original paper of \cite{genel_following_2013}: in addition to using a specific linked list for the tracer particles, the moving mesh of {\sc arepo} reduces the number of tracer movements and mitigates the cost of each tracer.

In comparison to the original paper by \cite{genel_following_2013}, we provide an additional detailed description of the statistical properties of the ensemble of tracer particles not only in the gas but also in stars and in AGN jets. We also studied how their distributions behave  when complex sub-grid models are involved (star formation and feedback, AGN feedback, BH accretion) and checked that their distribution is in agreement with the baryon distribution. 

This implementation provides an efficient method to accurately track the evolution of the Lagrangian history in the Eulerian code {\sc Ramses}. It opens new perspectives to study how baryon flows interact in hydrodynamical simulations. 
For instance, tracer particles could be used to quantify the spatial and time evolution of the anisotropically accreted gas, its contribution to the spin of  galaxies, and how these processes impact  galactic morphology.  Specifically, following \cite{tillson_angular_2015}, \cite{2015MNRAS.449.2087D}, and \cite{2017ApJ...841...16D}, one could address the following open questions: Where does the angular momentum go? Does it contribute to the spin-up of the galaxies or is it re-distributed before entering the disk? If it is, is it due to turbulent pressure, shock-heating or SN and AGN feedback? 

\begin{acknowledgements}
We wish to thank J.~Blaizot, J.~Devriendt, R.~Teyssier and M.~Trebitsch for useful suggestions. CC wishes to acknowledge the valuable feedback provided by R.~Beckmann and P.~Mitchell. CC is sponsored by the {\sl Institut Lagrange de Paris} fellowship. This work has made use of the Horizon Cluster hosted by {\sl Institut d'Astrophysique de Paris}. We thank Stephane Rouberol for running smoothly this cluster for us. It has also made use of the Occigen Cluster hosted by the CINES on the A0040406955 GENCI grant. This work has extensively used {\sc yt}, the open-source analysis and visualisation toolkit. The source code of the new tracer particles is available upon request.
\end{acknowledgements}

%%%%%%%%%%%%%%%%%%%%%%%%%%%%%%%%%%%%%%%%%%%%%%%%%%

%%%%%%%%%%%%%%%%%%%% REFERENCES %%%%%%%%%%%%%%%%%%

\bibliographystyle{aa}
\bibliography{biblio}

% \appendix
\begin{appendix}
\section{Tracer particle algorithm}%
\label{sec:algo}

Let us describe here the pseudo-code underlying the tracer particle algorithm. The corresponding {\sc Fortran} code is available upon request.

\subsection{Gas to gas cells}%
\begin{figure}
  \centering
  \includegraphics[width=.333\columnwidth]{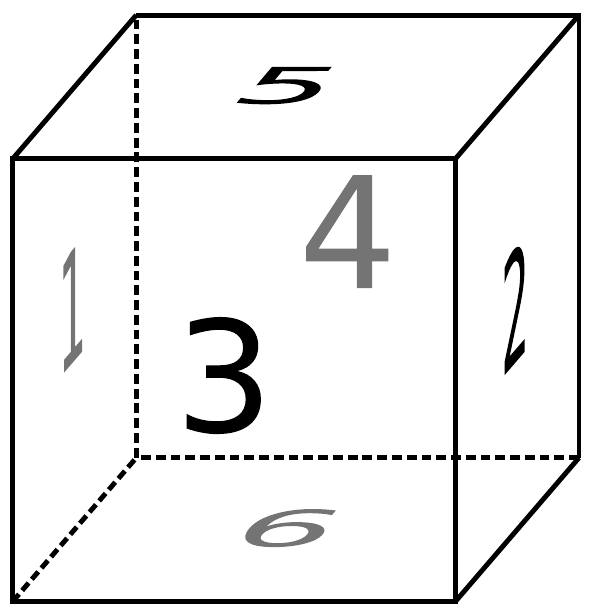}
  \caption{Cell faces numbering.}
  \label{fig:dice-scheme}
\end{figure}

\label{sec:algorithm-gasgas}
The main function in charge of moving tracers between gas cells is called \textsc{TreatCell}. It takes as input the index of a cell and loops over all tracers in it. It requires all the (mass) fluxes to be stored. The pseudo code is the following.
\begin{algorithmic}[5]
  \Function{TreatCell}{$i_\cell$}
    \State $m_\cell\gets$ \Call{MassOfCell}{$i_\cell$}
    \State $F_\text{net}\gets 0$
    \For{$i_\text{dir}\gets 1, 2N_\text{dim}$} \Comment{Compute outgoing flux}
      \State $F\gets$ \Call{GetFluxInDir}{$i_\cell$, $i_\text{dir}$}
      \If{$F > 0$}
         \State $F_\text{net} \gets F_\text{net}+F$
      \EndIf
    \EndFor

    \State $\text{tracers} \gets$ \Call{GetTracerParticlesInCell}{$i_\cell$}
    \State $p_\text{out}\gets F_\text{net}/m_\cell$ \Comment{Probability to move part. out of cell}

    \For{$j_\part$ {\bf in} tracers} \Comment{Loop on tracer particles}
      \State $r_1\gets$ \Call{DrawUniform}{$0$, $1$}
      \If{$r_1<p_\text{out}$}
         \State $r_2\gets$ \Call{DrawUniform}{$0$, $1$}
         \For{$i_\text{dir} \gets 1, 2N_\text{dim}$} \Comment{Select a direction}
           \State $F\gets$ \Call{GetFluxInDir}{$i_\cell$, $i_\text{dir}$}
           \State $p = F/F_\text{net}$
           \If{$r_2 < p$} \Comment{Move in direction $i_\text{dir}$}
             \State \Call{MoveParticle}{$i_\cell$, $j_\text{part}$, $i_\text{dir}$}
             \State \Break
           \Else
             \State $r_2\gets r_2 - p$ 
           \EndIf
         \EndFor
      \EndIf
    \EndFor
  \EndFunction
\end{algorithmic}
This function requires the \textsc{MoveParticle} function, which is defined as follow
\begin{algorithmic}[5]
  \Function{MoveParticle}{$i_\cell$, $i_\text{part}$, $i_\text{dir}$}
    \State $F_\text{tot}\gets$ \Call{GetFluxInDir}{$i_\cell$, $i_\text{dir}$}
    \State $\text{neighbors} \gets$ \Call{GetCellsOnFace}{$i_\cell$, $i_\text{dir}$}
    \State $\bar{i}_\text{dir} \gets$ \Call{GetOppositeDirection}{$i_\text{dir}$}
    \State $r\gets$ \Call{DrawUniform}{0, 1}
    \For{$j_\cell$ {\bf in} neighbors}
      \State $F\gets-$ \Call{GetFluxInDir}{$j_\cell$, $\bar{i}_\text{dir}$}
      \State $p\gets F/F_\text{tot}$
      \If{$r<p$}\Comment{Move particle to the centre of the cell}
        \State \Call{SetParticleAtCenter}{$i_\text{part}$, $j_\cell$}
        \State\Break
      \Else \Comment{Proceed to next cell}
        \State $r\gets r-p$
      \EndIf
    \EndFor
  \EndFunction
\end{algorithmic}
\textsc{GetFluxInDir} returns the mass that goes through the cell face in one timestep. Assuming that cell faces are numbered from 1 to 6 (left, right, top, bottom, front, rear, see Fig.~\ref{fig:dice-scheme}), \textsc{GetOppositeDirection} reads
\begin{algorithmic}[5]
  \Function{GetOppositeDirection}{$i_\text{dir}$}
  \State mask $\gets [2, 1, 4, 3, 6, 5]$
  \State {\bf return} $\mathrm{mask}[i_\text{dir}]$
  \EndFunction
\end{algorithmic}

When looped over all cells, the algorithm treating all the tracers has complexity $\mathcal{O}(N)$ where $N$ is the total number of tracer particles and requires $\mathcal{O}(N_\text{dim}N_\text{cell})$ memory to store the fluxes and $\mathcal{O}(N)$ to store the tracer particles information.

\subsection{AGN}

\label{sec:algorithm-agn}

Here we present how the volume of the jet is computed. We also present how the positions of the tracer particles in the jet are drawn. The function in charge of drawing position for the tracer particles in the jet is \textsc{Tracer2Jet}

\begin{algorithmic}[5]
  \Function{Tracer2Jet}{$\vec{j}$}
    \Loop
      \State $c\gets 2$
      \While{$c>1$}
        \State $a \gets$ \Call{NormalDistribution}{0, 1}
        \State $b \gets$ \Call{NormalDistribution}{0, 1}
        \State $c \gets a^2+b^2$
      \EndWhile
      \State $x\gets r_\AGN \times a$
      \State $y\gets r_\AGN \times b$
      \State $h\gets$ \Call{Uniform}{$-2r_\AGN, 2r_\AGN$}
      \State $r^2\gets x^2 + y^2$
      \If{$|h| > r_\AGN$ {\bf and} $(|h|-r_\AGN)^2+r^2<r_\AGN^2$}
        \State\Break
      \ElsIf{$|h| \leq r_\AGN$}
        \State \Break
      \EndIf
    \EndLoop
    \State\Comment{We now have a position in the frame of the jet.}
    \State $\vec{u}_z\gets \vec{j}/|\vec{j}|$
    \State $\vec{u}_x\gets [j_y+j_z,-j_x+j_z,-j_x-j_y]$
    \State $\vec{u}_x\gets \vec{u}_x / |\vec{u}_x|$
    \State $\vec{u}_y\gets \vec{u}_z\wedge \vec{u}_x$
    \State {\bf return} $x\ \vec{u}_x + y\ \vec{u}_y + h\ \vec{u}_z$
  \EndFunction  
\end{algorithmic}
\end{appendix}
% Don't change these lines
% \bsp{}% typesetting comment
\label{lastpage}
\end{document}